\documentclass[10pt]{iopart}

\usepackage[figuresright]{rotating} 
\newcommand{\eqref}[1]{(\ref{#1})}
\newcommand{\eqsref}[2]{(\ref{#1}--\ref{#2})}
\usepackage{bm,color}
\usepackage{hyperref}
\usepackage{mathrsfs}
\graphicspath{{./fig/}{./ps/}}
\usepackage{cite}

\def \ds{\displaystyle}
\def \pa{\partial}

\def\calEE{{\cal E}}
\def\xiM{\xi_{\rm M}}
\def\kf{k_\ast}

\newcommand{\Fig}[1]{Fig.~\ref{#1}}
\newcommand{\Figss}[2]{Figs.~\ref{#1}--\ref{#2}}
\newcommand{\Figp}[2]{Fig.~\ref{#1}(#2)}

\newcommand{\Sp}{{\rm{Sp}}}
\newcommand{\Eq}[1]{Eq.~(\ref{#1})}

\newcommand{\yjour}[5]{, ``#5,'' #2 {\bf #3}, #4 (#1).}
\begin{document}

\title[The SVT modes of GW turbulence simulations]{The scalar, vector, and tensor modes in gravitational wave turbulence simulations}

\author{Axel~Brandenburg$^{1,2,3,4}$, Grigol~Gogoberidze$^{4}$,  Tina~Kahniashvili$^{3,4,5}$, Sayan~Mandal$^{6,4}$\footnote{Corresponding author; all authors listed alphabetically.}, Alberto~Roper~Pol$^{7,4}$, Nakul~Shenoy$^{8}$}
\address{$^1$ Nordita, KTH Royal Institute of Technology and Stockholm University, Hannes Alfv\'ens v\"ag 12, 10691 Stockholm, Sweden}
\address{$^2$ Department of Astronomy, AlbaNova University Center, Stockholm University, 10691 Stockholm, Sweden}
\address{$^3$ Department of Physics and McWilliams Center for Cosmology, Carnegie Mellon University, 5000 Forbes Ave, Pittsburgh, PA 15213, USA}
\address{$^4$ School of Natural Sciences and Medicine, Ilia State University, 3-5 Cholokashvili Street, 0194 Tbilisi, Georgia}
\address{$^5$ Department of Physics, Laurentian University, Sudbury, ON P3E 2C, Canada}
\address{$^6$ Physics and Astronomy Department, Stony Brook University, 100 Nicolls Road, Stony Brook, New York 11794, USA}
\address{$^7$ Universit\'e de Paris, CNRS, Astrophysique et Cosmologie, Paris, F-75013, France}
\address{$^8$ Department of Physics, Massachusetts Institute of Technology, Cambridge, MA 02139, USA}

\ead{brandenb@nordita.org, grigol\_gogoberidze@iliauni.edu.ge, tinatin@andrew.cmu.edu, sayan.mandal@stonybrook.edu, roperpol@apc.in2p3.fr, getnakul@mit.edu (\today)}

%%%%%%%%%%%%%%%%%%%%%%%%%%%%%%%%%%%%%%%%%%
\begin{abstract} 
We study the gravitational wave (GW) signal sourced by primordial turbulence 
that is assumed to be present at cosmological phase transitions like 
the electroweak and quantum chromodynamics phase transitions.
We consider various models of primordial turbulence, such as those with and without helicity, purely 
hydrodynamical turbulence induced by fluid motions, and magnetohydrodynamic turbulence
whose energy can be dominated either by kinetic or magnetic energy, depending on the nature 
of the turbulence. 
We also study circularly polarized GWs generated by 
parity violating sources such as helical turbulence. 
Our ultimate goal is to determine the efficiency of GW production
through different  classes of turbulence. 
We find that the GW energy and strain tend to be large for acoustic or irrotational
turbulence, even though its tensor mode amplitude is relatively small at most wave numbers. 
Only at very small wave numbers is the spectral tensor mode significant,
which might explain the efficient GW production in that case.
\end{abstract}
%%%%%%%%%%%%%%%%%%%%%%%%%%%%%%%%%%%%%%%%%%%%%%%%%%


\maketitle

\section{Introduction}
\label{intro}

Gravitational waves (GWs) can provide a window into the earliest epochs of the universe when it
was opaque to electromagnetic radiation (see Ref.~\cite{Maggiore:2018sht}, Chapters: 21--22). 
Being massless and decoupled from other physical interactions (weak,
strong, and electromagnetic) GWs propagate 
{\it almost}\footnote{We discard the damping due to neutrino free streaming \cite{Durrer:1997ta,Weinberg:2003ur,Mangilli:2008bw,Lattanzi:2010gn,Stefanek:2012hj,Liu:2015psa,Saikawa:2018rcs}
or from anisotropic stresses \cite{Deryagin:1987ab}}
freely after their generation
preserving their spectral characteristics and diluting only with the
expansion of the universe. 
In particular, cosmological GW backgrounds produced by magnetohydrodynamic (MHD) turbulence
conserve the characteristics of the source, i.e., the spectral peak and the energy density of the
produced GW signal can be related to the characteristic scale and total energy density of the
sourcing magnetic or velocity field at the time of generation
(with a precision of about one order of magnitude for the energy density depending on
the duration of the turbulence sourcing) \cite{Kahniashvili:2020jgm}.

If GWs are generated from the early universe at energy scales 
around the 
electroweak (EW) and quantum chromodynamics (QCD) phase transitions are
detected, they will carry information about the fundamental physical 
processes at play when the universe was only $10^{-11}$ and $10^{-5}$ 
seconds old, respectively.
The Laser Interferometer Space Antenna (LISA), slated to launch in 2034,
will be sensitive to the GWs generated around the $100\,\mathrm{GeV}-1\,\mathrm{TeV}$ 
scale (see Refs.~\cite{Witten:1984rs,Hogan:1986qda,Kamionkowski:1993fg} for pioneering works and Ref.~\cite{Barausse:2020rsu} for
a review and reference therein).
It is also conceivable that the GWs generated at the EW epoch can be detected in
atomic interferometry experiments \cite{Dimopoulos:2007cj}, and those originated at the QCD epoch (or around)
at typical frequencies of $10^{-7}$--$10^{-9}$\,Hz 
in pulsar timing array (PTA) experiments
\cite{Caprini:2010xv,Ellis:2012in,Capozziello:2018qjs}.
Indeed PTAs provide detection of GW signals in the nHz regime making the GW signal arising at QCD 
energy scale accessible.
Recently, the NANOGrav collaboration has reported the detection of a signal, which could
be compatible with a stochastic GW background \cite{Arzoumanian:2020vkk}. 
If this detection is confirmed in the future, it
could be a manifestation of the stochastic GW background originated from the epoch around the QCD 
energy scale \cite{Ratzinger:2020koh,Neronov:2020qrl,Abe:2020sqb,Ramberg:2020oct,Kitajima:2020rpm,Li:2021qer,Gorghetto:2021fsn,Brandenburg:2021tmp,He:2021bqm,Garcia-Bellido:2021zgu}.
Interestingly, these GWs might be sourced by and reveal the presence of
primordial magnetic fields 
and turbulent motions 
at the QCD scale
\cite{Neronov:2020qrl,Brandenburg:2021tmp}.\footnote{The commonly discussed 
sources of GWs in the nHz regime are mergers of supermassive black holes 
\cite{Arzoumanian:2020vkk,Sesana:2012ak,Middleton:2020asl}.}.

Stochastic relic GW backgrounds can be generated 
through different 
violent
processes in the early universe, such as amplification of quantum-mechanical fluctuations during inflation, 
topological defects, cosmic strings, and cosmological phase transitions,
see Ref.~\cite{Caprini:2018mtu} for a review and references therein. 
In the case of cosmological phase transitions, GWs 
are generated 
either from the collision of bubbles 
during first order phase transitions, like the EW or QCD
phase transitions (see Refs.~\cite{Hogan:1986qda,Kamionkowski:1993fg} for pioneering works), 
or from the hydrodynamic and magnetohydrodynamic (MHD) turbulence
produced in association with the generation of primordial magnetic 
fields (PMFs) and turbulence 
at those early times; see  Refs.~\cite{Deryagin:1987ab,Durrer:1999bk,Kosowsky:2001xp} for pioneering works,
and for a recent review, see Sec.~8 of Ref~\cite{Caprini:2018mtu} 
and references therein. 

In this paper we focus on possible turbulent sources in the early 
universe and the formalism developed here is applicable independently 
of the nature of the origin of the turbulence. 
Additionally, our approach might be used when studying superfluid
turbulence in neutron stars and the GWs emitted by this process 
must be accounted for when estimating the continuous GW spectrum; 
see Ref.~\cite{Melatos:2009mz} for pioneering work and Sec.~6 of
Ref.~\cite{Lasky:2015uia} for a review and references therein.

The consideration of primordial magnetic fields (and correspondingly
turbulent motions) is motivated by observations of blazar spectra by 
the Fermi Gamma-ray Observatory.
They indicate the presence of magnetic fields at extragalactic
scales (see Ref.~\cite{Neronov:1900zz} for a pioneering work
and Refs.~\cite{Archambault:2017hvo,Biteau:2018tmv} for recent 
studies)\footnote{Also see Ref.~\cite{Arlen:2012iy} for
discussions on possible uncertainties in the measurements of blazar
spectra and Refs.~\cite{Broderick:2018nqf,AlvesBatista:2019ipr} on
possible impacts of plasma instabilities.}.
The high conductivity of the early universe plasma ensures strong 
coupling between magnetic fields and fluid motions, so the presence 
of primordial magnetic seeds inevitably results in the development of
turbulence \cite{Ahonen:1996nq,Brandenburg:1996fc}. 
This turbulence can also reinforce the magnetic field; see
Ref.~\cite{Brandenburg:2017rnt} for a discussion of the
dynamo mechanism in decaying turbulence. 
{\it Any other} turbulent processes in the early universe
can also amplify the magnetic field through dynamo action.
The stochastic magnetic fields and the turbulent velocity fields of the primordial 
plasma contribute to the anisotropic stress tensor that sources the primordial GWs. 
In addition there might be parity violating (or CP violating) processes in the early universe that lead to the generation of helical magnetic fields, or helical turbulent
motions, and thus produce circularly polarized GWs \cite{Kahniashvili:2005qi}.
In particular, the detection of circularly polarized GWs will shed light
on phenomena of fundamental symmetry breaking in the early universe,
such as parity violation, and potentially can serve as an explanation
of the lepto- and baryogenesis problem (see Refs.~\cite{Kuzmin:1985mm, Shaposhnikov:1986jp, Cohen:1990py, Cohen:1990it}
for pioneering works, and Ref.~\cite{Morrissey:2012db} for a recent review).

In the present work,
we study the scalar, vector, and tensor mode decomposition 
of different turbulence sources with the objective to help understand their 
efficiency in producing GWs. 
More precisely we have considered different classes of turbulence that
may be realized in the early universe.
In particular we have studied kinetically and magnetically dominant 
helical and non-helical cases as well as two types of acoustic turbulence.
The latter type consists of interacting sound waves.
At large enough Mach numbers, the flow is dominated by shocks.
Notably it was shown that the efficiency of sound waves as a source generating relic GWs
is not only comparable but exceeds that from bubble collisions; see Ref.~\cite{Hindmarsh:2013xza}
for a pioneering work, and Ref.~\cite{Hindmarsh:2019phv} for a recent study and references therein.
However the generation mechanism related to sound waves requires a first order PT. 
The ultimate goal of the work presented here is to determine the physical conditions in the
early universe that reach maximal efficiency of the GW signal production {\it without}
being limited by the first order PT consideration.
We also study helical vs non-helical sources behavior, and we confirm that,
for distinguishing between parity conserving and parity violating 
sources, the measurement of polarization degrees in the low frequency 
tail of the spectrum could be used to identify 
the inverse cascade (which applies to the helical turbulence case)
versus the slow inverse transfer. 

This paper is arranged as follows.
In Sec.~\ref{secBackg}, we briefly review the decomposition of the MHD stress tensor 
sourcing the GWs into the scalar, vector, and tensor components,
and provide a background on the modeling of turbulent sources.
This is followed by detailed numerical modeling for several types of
turbulent hydrodynamic and/or MHD flows in Sec.~\ref{secNumerics}.
In Sec.~\ref{secDegPol}, we describe in brief the degree of circular polarization
of both the source and the generated GWs, and present numerical simulations of
the degree of polarization of the source. 
Finally in Sec.~\ref{secConcl}, we present our conclusions.
Throughout this work, we set $\hbar=c=k_B=1$, and use the metric signature $(-,+,+,+)$.
We also set the permeability of free space to unity, i.e., $\mu_0=1$,
expressing the electromagnetic quantities in Lorentz-Heaviside units.
Repeated indices denote a summation, unless otherwise specified.
Greek indices run over spacetime coordinates, while Latin indices run over spatial coordinates.

\section{Formalism}
\label{secBackg}

The stress tensor that sources GWs in the early
universe receives contributions from PMFs \cite{Deryagin:1987ab}, as well as the
turbulent velocity field of the primordial plasma \cite{Christensson:2000sp, Kahniashvili:2010gp},
and the dynamics of phase transitions \cite{Witten:1984rs,Hogan:1986qda, Kamionkowski:1993fg}.
In the following subsection, we discuss the decomposition of the stress tensor into its scalar,
vector, and tensor (i.e., the transverse-traceless) components\footnote{This is not to
be conflated with the transverse-traceless gauge in which the gravitational waves are
computed far from the source.}, following the prescription by 
Lifshitz \cite{Lifshitz:1945du}, and in the next subsection,
we will present a brief overview of the different types of turbulent sources.
Readers who are familiar with the mechanism of SVT decomposition can skip ahead to Sec. \ref{secTurbMod}.

\subsection{The scalar-vector-tensor decomposition of the stress tensor}

\label{secSVT}

According to the Lifshitz prescription, any symmetric rank-2 tensor
$\lambda_{ij}(\mathbf{x})$ in $d$-dimensions can be decomposed as
\cite{Mukhanov:1990me}
\begin{equation}\label{eSVTMain}
\lambda_{ij}=L\delta_{ij}+\nabla_{\langle i}\nabla_{j\rangle}\lambda+\nabla_{(i}\bar{\lambda}_{j)}+\bar{\lambda}_{ij},
\end{equation}
where $\bar{\lambda}_i$ is divergenceless, $\nabla_i\bar{\lambda}_i=0$, $\bar{\lambda}_{ij}$ is divergenceless and traceless,
\begin{equation}\label{eBarLamPropMain}
\nabla_i\bar{\lambda}_{ij}=\nabla_j\bar{\lambda}_{ij}=0,\quad \bar{\lambda}_{ii}=0,
\end{equation}
and 
\begin{eqnarray}\label{eLamijScVecPartsMain}
\nabla_{\langle i}\nabla_{j\rangle}\lambda&\equiv\left(\nabla_i\nabla_j-\frac{1}{d}\delta_{ij}\nabla^2\right)\lambda,\\
\nabla_{(i}\bar{\lambda}_{j)}&\equiv
\frac{1}{2}\big(\nabla_i\bar{\lambda}_j+\nabla_j\bar{\lambda}_i\big),
\end{eqnarray}
correspond to the removal of trace, and the symmetrization operators,
respectively.
This decomposition is commonly referred to as a scalar-vector-tensor (SVT) decomposition.
Extracting the components in real space is algebraically involved, and is
discussed in detail in \ref{AppHelm}.

The process, however, is less cumbersome in momentum space where
the analog of Eq.~\eqref{eSVTMain} can be written as,
\begin{equation}\label{eLambSVTKSpace}
\tilde{\lambda}_{ij}= \tilde{L}\delta_{ij}-k_{\langle i}k_{j\rangle}\tilde{\lambda}^S+ik_{(i}\tilde{\lambda}^V_{j)}+\tilde{\lambda}^T_{ij},
\end{equation}
where tildes denote quantities in Fourier space\footnote{We also adopt
the notational convention that a function $f(\mathbf{r})$ of real space
coordinates $\mathbf{r}$ is expressed in terms of its Fourier transform
$\tilde{f}(\mathbf{k})$ as $f(\mathbf{r})=(2\pi)^{-3}\int d^3\mathbf{k}
\,e^{-i\mathbf{k}\cdot\mathbf{r}}\tilde{f}(\mathbf{k})$,
and the inverse transform is defined accordingly.}, $k_i$ is the $i$-th component
of $\mathbf{k}$, and the superscripts $S$, $V$, and $T$ denote the scalar, vector, and tensor
contributions respectively.
Analogous to the real space expressions, we have $k_i \tilde{\lambda}^V_i=0,
k_i \tilde{\lambda}^T_{ij}=k_j \tilde{\lambda}^T_{ij}=0$, and
$\tilde{\lambda}^T_{ii}=0$,  $k_{\langle i}k_{j\rangle}\equiv k_i 
k_j-k^2\delta_{ij}/d$
with $k=|\mathbf{k}|$, and the parentheses in the indices denote a symmetrization
as in Eq.~\eqref{eLamijScVecPartsMain}.
$\tilde{L}$ is proportional to the trace, i.e., $\tilde{L}=\tilde{\lambda}_{ii}/d$.
The other scalar part is extracted out as
\begin{equation}\label{eLKSpace} 
\tilde{\lambda}^S = -\frac{d}{d-1}\frac{\hat{k}_i\hat{k}_j}{k^2} \tilde{\lambda}_{\langle ij\rangle} 
= -\frac{d}{d - 1} \frac{\hat{k}_i \hat{k}_j}{k^2}
\left(\tilde{\lambda}_{ij} - \frac{1}{d} \delta_{ij} \tilde{\lambda}_{ii} \right),
\end{equation}
where $\hat{k}_i\equiv k_i/k$, and the vector contribution turns out to be 
\begin{eqnarray}\label{eLIKSpace} 
\tilde{\lambda}^V_i&=-\frac{2i}{k^2}\left(k_j\tilde{\lambda}_{\langle ij\rangle}+\frac{d-1}{d}k_{i}k^2\tilde{\lambda}^S\right) \nonumber\\
&=-\frac{2i}{k^2}\left(k_j\tilde{\lambda}_{\langle ij\rangle}-k_{i}\hat{k}_p\hat{k}_q\tilde{\lambda}_{\langle pq\rangle}\right).
\end{eqnarray} 
Finally, the tensor part can thus be extracted from Eq.~(\ref{eLambSVTKSpace}) as
\begin{eqnarray}\label{eLIJKSpace} 
\tilde{\lambda}^T_{ij}&=\tilde{\lambda}_{\langle ij\rangle} -\frac{d}{d-1}\hat{k}_{\langle i} \hat{k}_{j\rangle}\hat{k}_p\hat{k}_q\tilde{\lambda}_{\langle pq\rangle} \nonumber \\
&-\left[\left\{\hat{k}_i \hat{k}_p\tilde{\lambda}_{\langle jp\rangle}+ (i\longleftrightarrow j)\right\}-2\hat{k}_i\hat{k}_j\hat{k}_r\hat{k}_s\tilde{\lambda}_{\langle rs\rangle}\right].
\end{eqnarray}

The more conventional way to extract the transverse-traceless part of the
stress tensor is through the four-point projection operator $P_{ijlm}(\hat{k})$,
defined as $P_{ijlm}(\hat{k})\equiv \left[P_{il}P_{jm}-\frac{1}{2}P_{ij}P_{lm}\right](\hat{k})$
with $P_{ij}(\hat{k})=\delta_{ij}-\hat{k}_i\hat{k}_j$,
as is done, for example, in Ref.~\cite{Durrer:1997ep} (see their Eq.~(22)).
It can be shown in a straightforward way that the result of this operation
gives the same expression as in Eq.~\eqref{eLIJKSpace}.
We now apply this formalism to the energy-momentum tensor for 
the magnetic 
and the plasma velocity fields.

\subsection{SVT Decomposition of Magnetic Stress}
\label{SVTDecompMagneticStress}

For a magnetic field $\mathbf{B}$, the stress tensor can be written as
\begin{equation}\label{eStressFirst}
\tau_{ij}(\mathbf{x})=B_i(\mathbf{x}) B_j(\mathbf{x})-\frac{1}{2}\delta_{ij}B^2(\mathbf{x}),
\end{equation}
where $B_i$ are the spatial components of the three-dimensional vector $\mathbf{B}$.
The magnetic field $\mathbf{B}$ is solenoidal, i.e., $\nabla\cdot\mathbf{B}=0$. 
This tensor can be decomposed into its SVT contributions using Eq.~\eqref{eSVTMain}, 
\begin{eqnarray}\label{eTauSVT} 
\tau_{ij}&=T\delta_{ij}+\nabla_{\langle i}\nabla_{j\rangle}\tau+\nabla_{(i}\bar{\tau}_{j)}+\bar{\tau}_{ij}, 
\end{eqnarray}
where $\nabla_i\bar{\tau}_i=0$, $\nabla_i\bar{\tau}_{ij}=\nabla_j\bar{\tau}_{ij}=0$, and $\bar{\tau}_{ii}=0$. 
An example is given in \ref{Examples} for a one-dimensional Beltrami field.
The term proportional to the trace is $T=-B_i B_i/6 \equiv -B^2/6$, and the other
scalar component is
(see \ref{AppHelm} for the derivation of these expressions for general
$\lambda_{ij}$)
\begin{equation}\label{eTau}
\tau=-\frac{1}{2}\int dV' \int dV''\frac{\ds{\nabla''^2\big[B^2(\mathbf{x}'')\big]}-3\nabla''_j B_i(\mathbf{x}'')\nabla''_i B_j(\mathbf{x}'')}{16\pi^2 |\mathbf{x}-\mathbf{x}'| |\mathbf{x}'-\mathbf{x}''|},
\label{tau_magnetic}
\end{equation}
and the vector contribution is
\begin{equation}\label{eBarTauIMag}
\bar{\tau}_j=\int dV' \,\frac{\ds{\nabla'_j\big\{B^2(\mathbf{x}')}+2\nabla'^2\tau(\mathbf{x}')\big\}-3B_i(\mathbf{x}')\nabla'_i B_j(\mathbf{x}')}{6\pi|\mathbf{x}-\mathbf{x}'|}.
\label{tauj_magnetic}
\end{equation}
Note that the solenoidality of the magnetic field has been used to obtain Eqs.~\eqsref{tau_magnetic}{tauj_magnetic}.
The tensor mode is then easily extracted by subtracting the scalar and vector contributions,
\begin{equation}\label{eBarTauIJ}
\bar{\tau}_{ij}=B_{\langle i}B_{j\rangle}-\nabla_{\langle i}\nabla_{j\rangle}\tau-\nabla_{(i}\bar{\tau}_{j)},
\end{equation}
where, as before, $B_{\langle i}B_{j\rangle}\equiv\left(B_i B_j-\delta_{ij}B^2/3\right)$.
In Fourier space, the stress tensor in Eq.~\eqref{eStressFirst} becomes a
convolution, 
\begin{eqnarray}\label{eStressFirstKSpace} 
\tilde{\tau}_{ij}(\mathbf{k})&=\frac{1}{(2\pi)^3}\left(\tilde{B}_i\star\tilde{B}_j\right)(\mathbf{k})\equiv\frac{1}{(2\pi)^3}\int d^3\mathbf{p}\,\tilde{B}_i(\mathbf{p}) \tilde{B}_j(\mathbf{k}-\mathbf{p}), 
\end{eqnarray}
where $\tilde{B}_i(\mathbf{k})$ is the Fourier transform of $B_i(\mathbf{x})$,
and we have omitted the isotropic term in Eq.~\eqref{eStressFirst}.
For the stress-tensor in Eq.~\eqref{eTauSVT}, the expression analogous to
Eq.~\eqref{eLambSVTKSpace} is,
\begin{equation}\label{eTauSVTKSpace}
\tilde{\tau}_{ij}=\tilde{T}\delta_{ij}-k_{\langle i}k_{j\rangle}\tilde{\tau}^S+ik_{(i}\tilde{\tau}^V_{j)}+\tilde{\tau}^T_{ij},
\end{equation} 
with $k_i \tilde{\tau}^V_i=0, k_i \tilde{\tau}^T_{ij}=k_j \tilde{\tau}^T_{ij}=0$, and $\tilde{\tau}^T_{ii}=0$.
The term proportional to the trace is 
$\tilde{T}=(2\pi)^{-3}(\tilde{B}_s\star\tilde{B}_s)(\mathbf{k})/3$, while
\begin{equation}\label{eTKSpace} 
\tilde{\tau}^S(\mathbf{k})=-\frac{1}{(2\pi)^{3}}\frac{3}{2}\frac{\hat{k}_i\hat{k}_j}{k^2}\tilde{B}_{\langle i}\star \tilde{B}_{j\rangle}(\mathbf{k}),
\end{equation}
where, continuing the spirit of using angle bracket notations,
$\tilde{B}_{\langle i}\star\tilde{B}_{j\rangle}(\mathbf{k})\equiv (\tilde{B}_i\star\tilde{B}_j)(\mathbf{k})-\delta_{ij}(\tilde{B}_s\star\tilde{B}_s)(\mathbf{k})/3$.
For clarity, we suppress the functional dependence on $\mathbf{k}$ henceforth.
The vector contribution turns out to be
\begin{equation}\label{eTIKSpace}
\tilde{\tau}^V_i=-\frac{2 i k_j}{k^2}\left[\frac{1}{(2\pi)^{3}}\tilde{B}_{\langle i}\star \tilde{B}_{j\rangle}+k_{\langle i}k_{j\rangle}\tilde{\tau}^S\right],
\end{equation}
and finally, the tensor part can thus be extracted as
\begin{equation}\label{eTIJKSpace}
\tilde{\tau}^T_{ij}=\frac{1}{(2\pi)^{3}}\tilde{B}_{\langle i}\star \tilde{B}_{j\rangle}+k_{\langle i}k_{j\rangle}\tilde{\tau}^S-ik_{(i}\tilde{\tau}^V_{j)}.
\end{equation}

\subsection{SVT Decomposition of Kinetic Stress}

For a fluid with velocity $\mathbf{u}$, the stress tensor analogous
to Eq.~\eqref{eStressFirst} can be written as
\begin{equation}\label{eStressFirstVel}
\sigma_{ij}(\mathbf{x})=w \gamma^2 u_i u_j-p\delta_{ij},
\end{equation}
where $w=p+\rho$ is the enthalpy\footnote{In a magnetized plasma, an important quantity associated 
with the magnetic field is the \textit{Alfv\'{e}n velocity} $\mathbf{v}_{\rm A}$ defined as 
$\mathbf{v}_{\rm A} =\mathbf{B}/\sqrt{w}$.},
$p$ and $\rho$ are the pressure and density of the fluid respectively,
$\gamma = (1 - u^2)^{-1/2}$ is the Lorentz factor,
and the spacetime dependence of all the fields are suppressed\footnote{If we have a
solenoidal flow, i.e., $\nabla\cdot\mathbf{u}=0$, then the anisotropic stress tensor for the
fluid velocity $\gamma u_i$ is analogous to that of magnetic fields.}.
The SVT decomposition of this tensor, analogous to Eq.~\eqref{eTauSVT}, can be written as
\begin{equation}\label{eSigmaSVT}
\sigma_{ij}=S\delta_{ij}+\nabla_{\langle i}\nabla_{j\rangle}\sigma+\nabla_{(i}\bar{\sigma}_{j)}+\bar{\sigma}_{ij},
\end{equation}
and as before, $\nabla_i\bar{\sigma}_i=0$, $\nabla_i\bar{\sigma}_{ij}=\nabla_j\bar{\sigma}_{ij}=0$, and $\bar{\sigma}_{ii}=0$.
The mathematical details are more involved compared to the case of magnetic fields,
since the velocity fields are allowed to have a longitudinal component,
as opposed to the case with magnetic fields.
In addition, we consider non-relativistic fluid velocities ($\gamma \sim 1)$ for simplicity of the expressions in the analytical results.
However, this term is recovered in the numerical treatment. 
We omit the general expressions for the S, V, and T components
of the stress tensor in real space,
which are provided in \ref{AppSVTFluid}.

The stress tensor can be written in $k$-space
in a manner analogous to Eq.~\eqref{eStressFirstKSpace} as 
\begin{eqnarray}\label{eStressFirstVelKSpace}
\tilde{\sigma}_{ij}(\mathbf{k})&=\frac{1}{(2\pi)^{6}}\left((\tilde{p}+\tilde{\rho})\star\tilde{u}_i\star\tilde{u}_j\right)(\mathbf{k}) -\tilde{p}(\mathbf{k})\delta_{ij} \nonumber \\
&=\frac{1}{(2\pi)^{6}}\int d^3\mathbf{p} \int d^3\mathbf{q}\,\bigg[(\tilde{p}+\tilde{\rho})(\mathbf{p})\tilde{u}_i(\mathbf{q}) 
\tilde{u}_j(\mathbf{k}-\mathbf{p}-\mathbf{q})\bigg] \nonumber\\
&-\tilde{p}(\mathbf{k})\delta_{ij}.
\end{eqnarray}
This is again decomposed as,
\begin{equation}\label{eSigmaSVTKSpace}
\tilde{\sigma}_{ij}=\tilde{S}\delta_{ij}-k_{\langle i}k_{j\rangle}\tilde{\sigma}^S+ik_{(i}\tilde{\sigma}^V_{j)}+\tilde{\sigma}^T_{ij},
\end{equation}
where $k_i \tilde{\sigma}^V_i=0, k_i \tilde{\sigma}^T_{ij}=0$, and $\tilde{\sigma}^T_{ii}=0$.
$\tilde{S}$ is again proportional to the trace,
$\tilde{S}=-\tilde{p}(\mathbf{k})+(2\pi)^{-6}((\tilde{p}+\tilde{\rho})\star\tilde{u}_i\star\tilde{u}_i)(\mathbf{k})/3$,
and the other scalar part is
\begin{equation}\label{eSKSpace} 
\tilde{\sigma}^S(\mathbf{k})=-\frac{1}{(2\pi)^6}\frac{3}{2}\frac{\hat{k}_i\hat{k}_j}
{k^2}(\tilde{p}+\tilde{\rho})\star \tilde{u}_{\langle i}\star \tilde{u}_{j\rangle}(\mathbf{k}),
\end{equation}
and as in Eqs.~\eqref{eTIKSpace} and \eqref{eTIJKSpace}, we have
\begin{equation}\label{eSIKSpace}
\tilde{\sigma}^V_i=-\frac{2 i k_j}{k^2}\left[\frac{1}{(2\pi)^6}(\tilde{p}+\tilde{\rho})\star\tilde{u}_{\langle i}\star \tilde{u}_{j\rangle}+k_{\langle i}k_{j\rangle}\tilde{\sigma}^S\right],
\end{equation}
and,
\begin{equation}\label{eSIJKSpace}
\tilde{\sigma}^T_{ij}=\frac{1}{(2\pi)^3}(\tilde{p}+\tilde{\rho})\star\tilde{u}_{\langle i}\star \tilde{u}_{j\rangle}+k_{\langle i}k_{j\rangle}\tilde{\sigma}^S-ik_{(i}\tilde{S}^V_{j)}.
\end{equation}

\subsection{Modeling of Turbulent Sources}
\label{secTurbMod}

As mentioned above, the primordial plasma that generates the GWs is highly turbulent,
and contains magnetic fields.
It is therefore essential to study the MHD evolution of the plasma and understand
the various kinds of turbulent initial conditions.
The purpose of this subsection is to provide an elementary background of MHD turbulence,
in preparation for the numerical results discussed in Sec.~\ref{secNumerics}.

The primordial plasma is described in a spatially flat, expanding, isotropic, 
and homogeneous universe, characterized by the 
Friedmann-Lema\^{i}tre-Robertson-Walker (FLRW) metric tensor,
$ds^2=-a^2(dt^2+ \delta_{ij}\,dx_i dx_j),$
where $\mathbf{x}$ are the comoving coordinates, $t$ is the conformal time,
and  $a$ is the scale factor.
We consider turbulent sources with subrelativistic bulk velocities, 
such that the Lorentz factor can be expanded up to second order,
$\gamma^2 \sim 1 + (v/c)^2$, 
and use the relativistic equation of state, $p=\rho c^2/3$, which is appropriate 
to describe a plasma dominated by massless particles.
The resulting MHD equations are similar to 
those in flat spacetime after rescaling the magnetic fields, the energy,
density and the spacetime coordinates to comoving variables and
conformal time \cite{Brandenburg:1996fc}.
We consider the energy density in the turbulence to be below 10\% of the
total energy density, which is consistent with the assumption of
subrelativistic bulk velocities. 
The upper limit for a magnetic energy density of 10\% of the energy
densities of additional relativistic components is required based on
big bang nucleosynthesis (BBN) bounds on the observed abundance of light
elements \cite{Grasso:1996kk}. 
On the other hand, for kinetic turbulence, we defer the extension
to relativistic velocities to future work.
It should be said, however, that important phenomena in MHD turbulence
such as inverse cascading are similarly reproduced both in relativistic
\cite{Zrake:2014mta} and nonrelativistic \cite{Brandenburg:2014mwa} simulations. 
Regarding GW production, the assumption of subrelativistic
bulk velocities is justified, since it has been shown
that neglecting the effect of thermal dissipation induced by shocks in the
highly relativistic regime leads to an underestimation of the resulting
GW signal; see Ref.~\cite{Kosowsky:2001xp}.

Turbulent flows are characterized by the magnetic and fluid Reynolds numbers,
$\mathrm{Re}_M$ and $\mathrm{Re}$ respectively, defined as
\begin{equation}\label{eRey}
\mathrm{Re}_M\equiv\frac{vL}{\eta},\quad\quad \mathrm{Re}\equiv\frac{vL}{\nu},
\end{equation}
for a typical fluid velocity $v$ on the length scale $L$, and $\eta$ and
$\nu$ being the magnetic diffusivity and kinematic viscosity respectively.
$\mathrm{Re}_M$ characterizes the relative importance of the diffusion term
over the induction term, while $\mathrm{Re}$ tells us
about the importance of the nonlinear velocity term
relative to the viscous dissipation term.

In the early universe, the viscosity and magnetic diffusivity are very small,
leading to a very large value of the Reynolds numbers \cite{Ahonen:1996nq}. 
Under such conditions, numerical MHD simulations in an expanding universe
have demonstrated the occurrence of a direct cascade 
from larger to smaller scales, indicating the presence of turbulence;
see Ref.~\cite{Brandenburg:1996fc} for pioneering work.

In MHD simulations, one generally studies the evolution of the energy
spectra 
$E_i(k)$ of the velocity and magnetic fields, for $i=K,M$ respectively 
(these energy spectra are defined such that the mean
energy density $\mathcal{E}_i$ is given by $\mathcal{E}_i=\int dk\,E_i(k)$). 
In a turbulent medium, the energy spectra have a range of scales where
the spectral shape is given by the Kolmogorov spectrum, i.e., $E_i(k)\propto k^{-5/3}$ 
-- this is called the inertial subrange; see \cite{Brandenburg:2009tf} for a review.
Scales smaller than the inertial range are dominated by viscous eddies,
while larger scales have energy carrying eddies.
Turbulence involves nonlinear dynamics, which leads to an energy cascade
whereby energy is transferred from larger to smaller scales (direct cascade)
or vice versa (inverse cascade).
The latter, for example, is interesting in the context of evolution
of primordial magnetic fields, since it can explain the presence of
magnetic fields correlated at sufficiently large scales in the universe
after they evolve through MHD evolution of the plasma.

The physics governing the MHD evolution of the magnetic and velocity fields
of the plasma affects the instantaneous decay exponents of the mean energies.
For example, if we have helical magnetic fields in a highly conducting plasma,
magnetic helicity is conserved throughout the evolution; in that case,
magnetic energy decays with time $t$ as $\mathcal{E}_M(t)\propto t^{-2/3}$.
On the other hand, for non-helical fields, one has $\mathcal{E}_M(t)\propto t^{-1}$. 
The correlation length of the turbulence, $\xiM$, which
is defined as a weighted average over the spectrum as
$\xiM=\mathcal{E}_i^{-1}\int dk\,k^{-1} E_i(k)$,
tends to increase in a corresponding fashion.
The increase is fastest for helical turbulence, where $\xiM\propto t^{2/3}$,
and slower for nonhelical turbulence, where $\xiM\propto t^{1/2}$ \cite{Brandenburg:2016odr}.
This can have a noticeable effect on the resulting GW production
\cite{Kahniashvili:2020jgm}.

\section{Numerical scalar--vector--tensor decomposition of the stress tensor}
\label{secNumerics}

\subsection{Numerical modeling setup}

In the following, we present the results of a numerical SVT decomposition 
of the stress tensor.
We consider four distinct types of isotropic hydrodynamic and MHD
turbulence similar to those considered in recent numerical studies of
GW production \cite{Pol:2019yex,Kahniashvili:2020jgm}.
The goal of those studies was to determine the form and
amplitude of the GW spectra from numerical simulations of hydrodynamic
and MHD turbulence.
It was found that the efficiency of converting kinetic and magnetic
energies into GW energy can be very different for different types
of flows.
The question therefore arises, whether this difference can be related
to differences in the associated SVT decomposition.
These flows are distinguished by different types of forcing; a
summary is given in Table~\ref{Tsummary}.  

The forcing energy is injected at wave numbers around $\kf$, which we
arrange to be six times larger than the smallest wavenumber $k_1=2\pi/L$
in our triply-periodic computational domain of side length $L$.
This forcing term is added to model the kinetic energy injection
produced by different mechanisms that lead to turbulence generation during 
the early universe.
Some examples include the expansion and collision of bubbles during first-order
phase transitions \cite{Witten:1984rs} or sound 
wave production due to scalar field perturbations \cite{Hindmarsh:2013xza}. 
Additionally, we model magnetic field generation in a similar way by
injecting energy at a characteristic scale that depends on the magnetogenesis
scenario; see Ref.~\cite{Vachaspati:2020blt} for a recent review. 
Axion-driven magnetogenesis can also induce turbulence at the QCD scale,
and does not require the phase transition to be of first-order; see, e.g.,
Ref.~\cite{Miniati:2017kah}. 
Another possibility is to consider pre-existing (for example,
inflationary) magnetic fields \cite{Kahniashvili:2012vt}.
For most of the cases, we choose $k_1=100\,H_*$, and in two cases,
we use $k_1=0.3\,H_*$). 
Energy is dissipated through a Laplacian diffusion operator with
coefficients $\nu$ and $\eta$ for the viscosity and the magnetic
diffusivity, which act on the smallest resolved length scales.
We use conformal time normalized to the Hubble time at the time of source 
initiation $H_\star^{-1}$, which corresponds to $t=1$ in our units.
Following the recent work of Ref.~\cite{Kahniashvili:2020jgm}, we apply 
a forcing term (in the momentum equation or in the induction equation
for kinetic or magnetic turbulence, respectively) for one Hubble time until $t=2$. 
The forcing has an initial stage $1 < t < 2$ when the forcing term is present,
an intermediate stage $2 < t < 3$ when it decreases linearly, and a final stage
$t > 3$ when it is absent.
The forcing is random and applied at each time step.
This allows to sustain stationary turbulence for a finite time during the intermediate
stage, which leads to freely decaying turbulence in the final stage. 
For Run~A, we apply forcing in the form of spherical expansion waves
through the gradients of randomly placed Gaussians with amplitude $f_0$
\cite{Mee:2006mq}, which are irrotational, i.e., curl-free.
This case was also studied in Ref.~\cite{Pol:2019yex}, and is similar
to their Run~ac1, except that in their case the forcing was only 
applied until $t=1.1$, and the intermediate stage is not considered.
In Run~B, we also apply irrotational forcing, but in the form of plane waves,
again with amplitude $f_0$.
The remaining four runs also have plane wave forcings, but are all vortical,
calculated as a curl of a randomly oriented vector with amplitude $f_0$ and
can have helicity, which is quantified by the parameter $\sigma \in [0, 1]$.
Run~C has nonhelical forcing, while Run~D has a helical forcing.
Runs~E and F have magnetic fields and the forcing is applied in the
induction equation, again first without helicity (Run~E) and with helicity
(Run~F).
In Runs~G and H, forcing is applied at small $k$ ($k_*=2$) during a time
interval $1\leq t\leq2$, using either vortical (Run~G) or irrotational
(Run~H) forcing. 
In all cases, we present the results of the SVT decomposition in the
form of power spectra per linear wave number interval for the projected
part of the stress tensor. 
The efficiency of GW production will be related to the contribution by the tensor mode
with respect to the scalar and vector modes 
of the turbulent source. 
The energy injection wave numbers are $k=k_*\equiv600$ (Runs~A--F) and 2 (Runs~G and H). 
The resulting spectra of the stress peak at $k=2k_*\approx1200$ and 4, respectively. 
Note that we have used normalized wave numbers $k$ such that
$k = 1$ corresponds to the
Hubble scale $H_*^{-1}$ at the time of generation.

\subsection{Comparison for Acoustic and Vortical Turbulence}

In Table~\ref{Tsummary}, we list the maximum kinetic or magnetic energies
$\calEE_i^{\max}$, where $i={\rm Ka}$ and ${\rm Kv}$ refer to kinetic
energy densities for acoustic and vortical turbulence, and ${\rm Mv}$
for magnetic energy densities, where the turbulence is also vortical.
These values are normalized by the radiation energy, just like the resulting
GW energy, $\calEE_{\rm GW}^{\rm sat}$, which refers to the GW energy
averaged over oscillations around the saturated stationary value.
In earlier work of Ref.~\cite{Pol:2019yex}, the saturation GW energy
$\calEE_{\rm GW}^{\rm sat}$ was found to be proportional to the square
of the ratio $\calEE_{i}^{\rm max}/k_*$.
To characterize the efficiency in GW energy conversion,
we quote in Table~\ref{Tsummary} the factor
$q=k_*\,(\calEE_{\rm GW}^{\rm sat})^{1/2}/\calEE_i^{\max}$.
In Ref.~\cite{Pol:2019yex}, the values of $q$ were found to be
between one and ten, but here we find even larger values.

\begin{figure}[t!]\begin{center}
%cvs co sayan/GW/K512irro_k6_ramp1c  #(and likewise for other runs)
\includegraphics[width=.49\columnwidth]{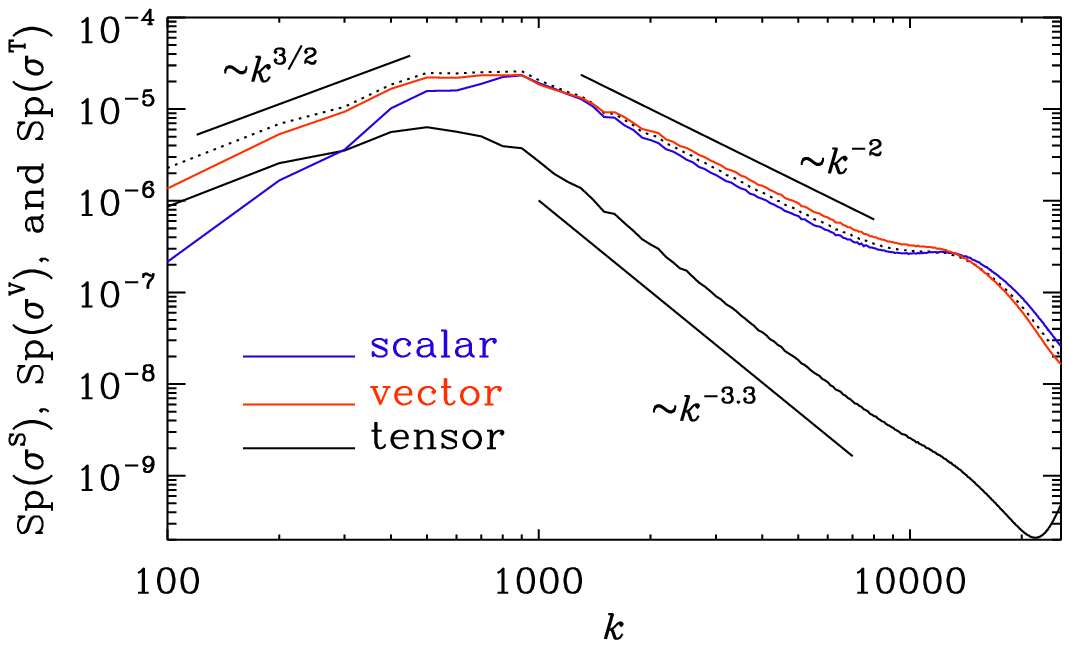}
\includegraphics[width=.49\columnwidth]{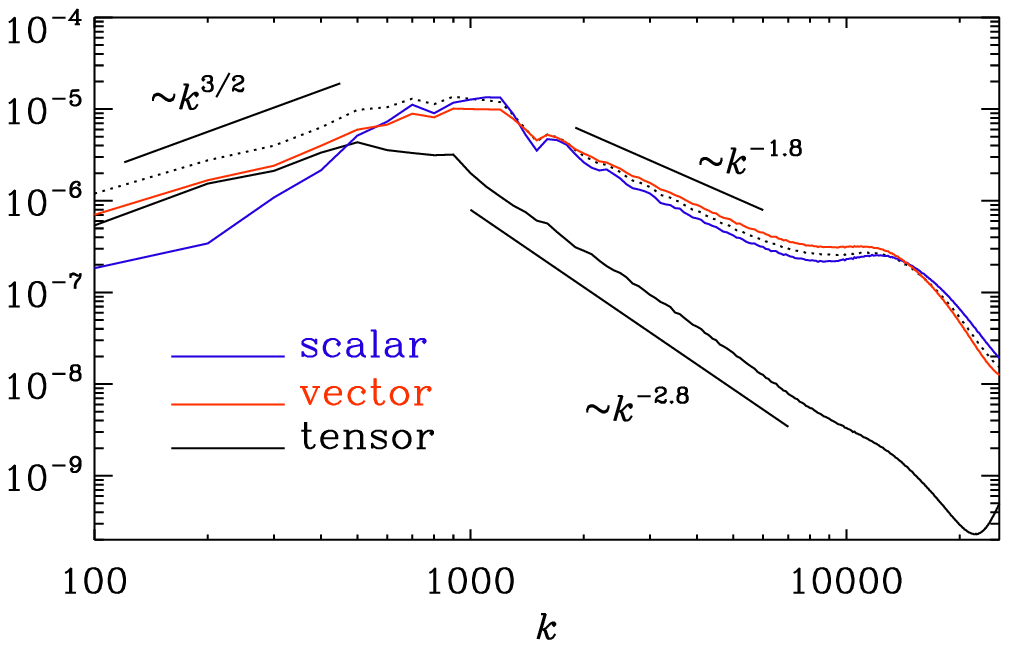}
\end{center}\caption[]{
SVT decomposition for Runs A (left) and B (right) for irrotational
turbulence produced by spherical expansion waves and plane wave forcing,
respectively.
}\label{pencil_ac}\end{figure}

We begin with the case of acoustic turbulence in Runs~A and B.
This type of turbulence is unusual in that it processes very little vorticity. 
In \Fig{pencil_ac}, we show the power spectra of the projected stress
tensor for acoustic turbulence. 
Those flows have always zero helicity.
The spectrum has a subinertial range proportional to $k^{3/2}$ 
and an inertial range which is proportional to $k^{-2}$
for Run~A and proportional to $k^{-1.8}$ for Run~B,
similar to what is expected for acoustic turbulence \cite{KP73}.
The scalar and vector components dominate over the tensor components at
almost all wave numbers, although the scalar component becomes smaller
than the tensor component at small wave numbers ($k<400$).
In addition, at small wave numbers, the vector components dominate over
the scalar component.

\begin{table}[t]\caption{
Summary of the runs discussed in the paper.
}\vspace{12pt}\centerline{\begin{tabular}{lccccccccc}
Run&$i$&$\kf$&$\calEE_i^{\max}$&$\sigma$&wave& $\calEE_{\rm GW}^{\rm sat}$ &$\nu$ ($=\eta$) & $f_0$ & $q$ \\
\hline
A & Ka &600&$3.8\times10^{-3}$&0&expan &$1.5\times10^{-8}$&$       10^{-6}$&$1             $ & 19 \\%K512irro_k6_ramp1c
B & Ka &600&$2.4\times10^{-3}$&0&plane &$1.6\times10^{-8}$&$2\times10^{-7}$&$2\times10^{-3}$ & 32 \\%K512irrok6c
C & Kv &600&$8.0\times10^{-3}$&0&plane &$4.5\times10^{-9}$&$2\times10^{-7}$&$4\times10^{-1}$ & 4.3 \\%K512sig0_k6_ramp1a
D & Kv &600&$1.1\times10^{-2}$&1&plane &$6.3\times10^{-9}$&$2\times10^{-7}$&$4\times10^{-1}$ & 5.0 \\%K512sig1_k6_ramp1a
E & Mv &600&$5.7\times10^{-3}$&0&plane &$1.5\times10^{-9}$&$2\times10^{-7}$&$6\times10^{-4}$ & 4.1 \\%M512sig0_k6_ramp1b
F & Mv &600&$1.7\times10^{-2}$&1&plane &$5.1\times10^{-9}$&$2\times10^{-7}$&$6\times10^{-4}$ & 2.5 \\%M512sig1_k6_ramp1b
G & Ka & 2 &$4.2\times10^{-2}$&0&plane &$8.3\times10^{-4}$&$2\times10^{-2}$&$7\times10^{-1}$ & 1.4 \\%K512irro_k6_wav03_7p0em1d_512procs
H & Kv & 2 &$4.7\times10^{-2}$&0&plane &$1.1\times10^{-3}$&$1\times10^{-2}$&$4\times10^{-1}$ & 1.4 \\%K512sig1_k6_wav03_4p0em1f
\label{Tsummary}\end{tabular}}\end{table}

\begin{figure}[t!]\begin{center}
\includegraphics[width=0.49\columnwidth]{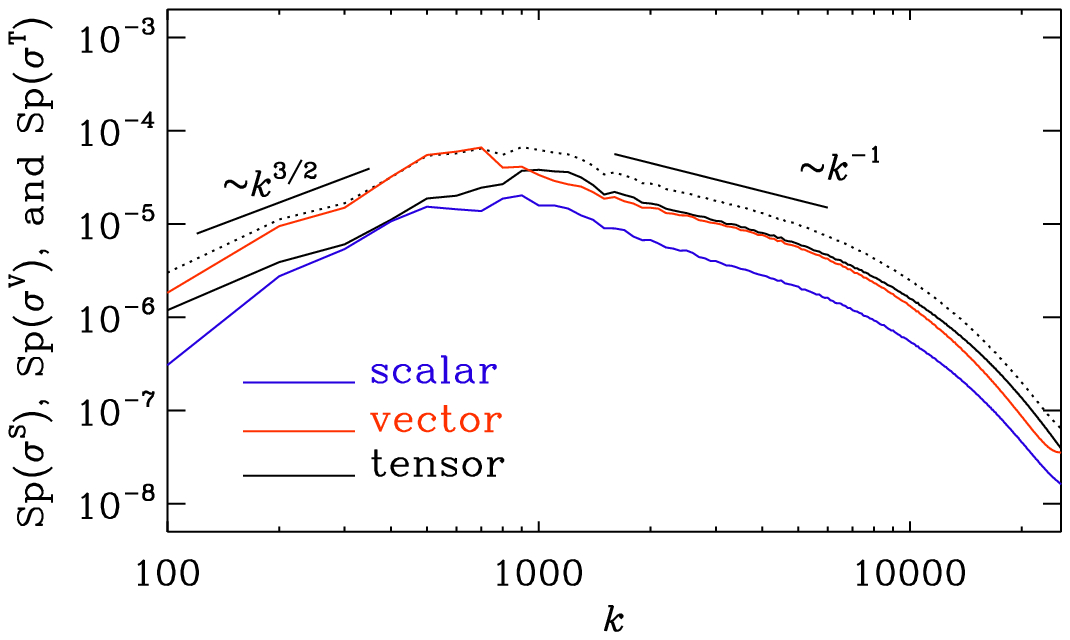}
\includegraphics[width=0.49\columnwidth]{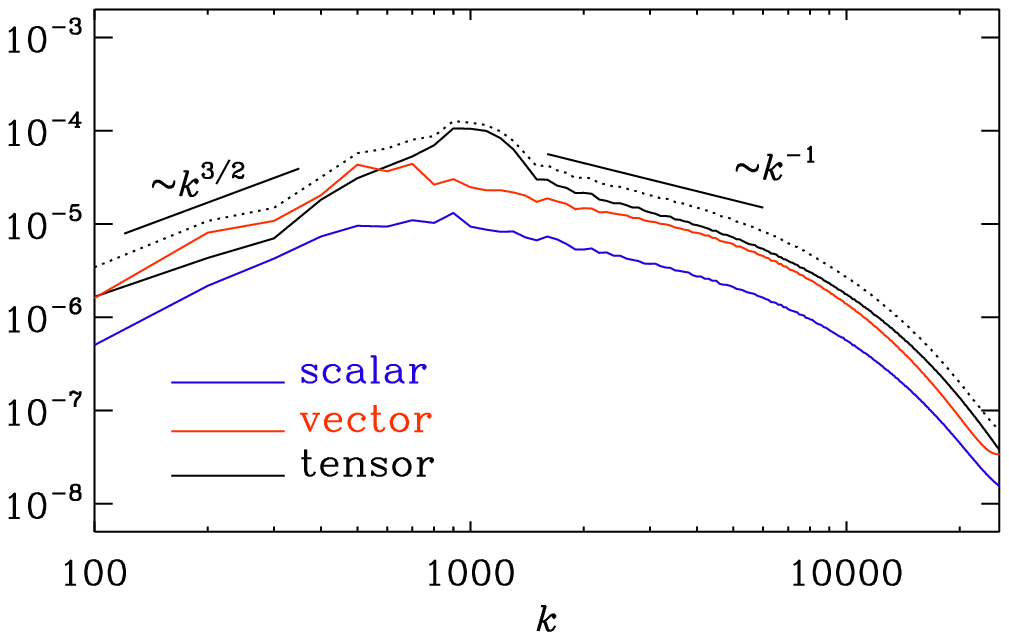}
\end{center}\caption[]{
SVT decomposition for Runs C (left) and D (right) for vortical
hydrodynamic turbulence without and with helicity, respectively. 
}\label{pencil_hyd}\end{figure}

\begin{figure}[t!]\begin{center}
\includegraphics[width=0.49\columnwidth]{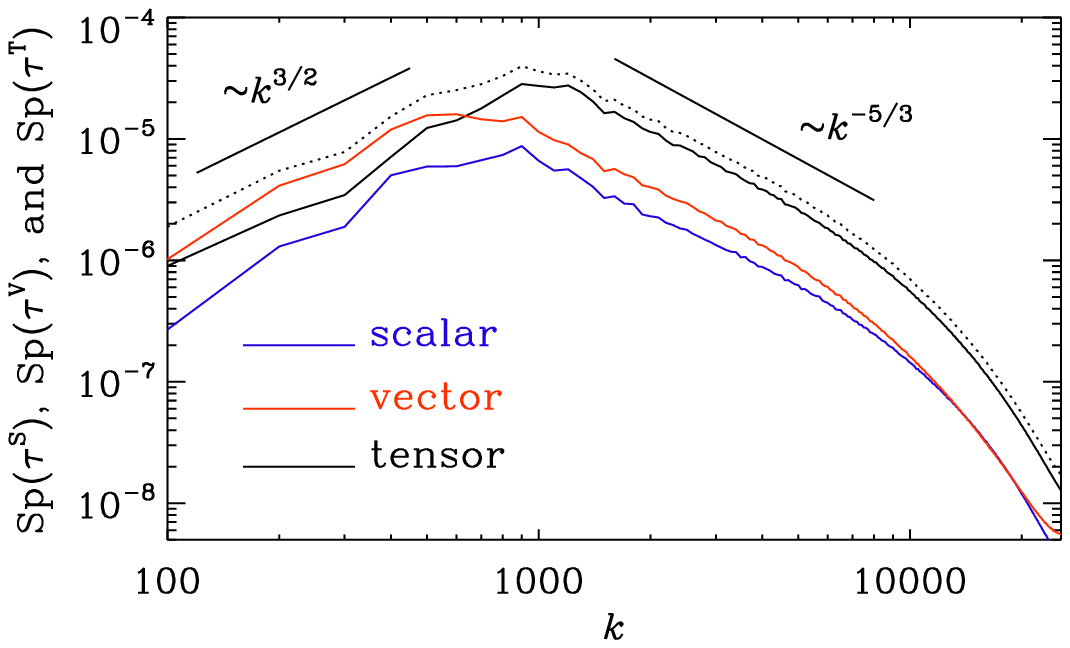}
\includegraphics[width=0.49\columnwidth]{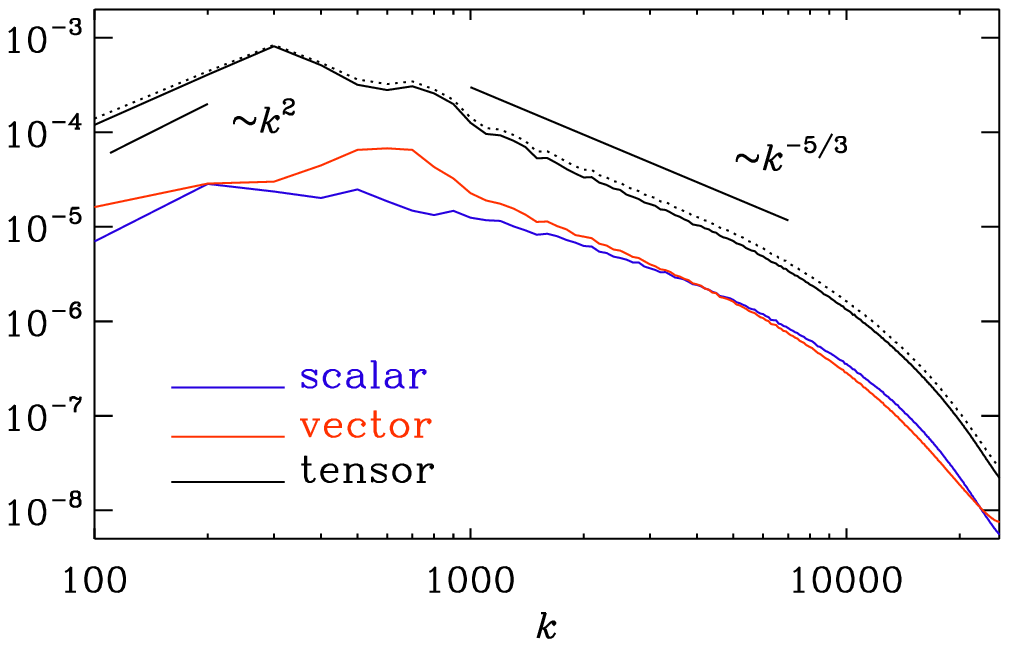}
\end{center}\caption[]{
SVT decomposition for Runs E (left) and F (right) with magnetic forcing
without and with helicity, respectively.
}\label{pencil_hel}\end{figure}

The situation is quite different for vortical turbulence.
In \Fig{pencil_hyd} we show cases without kinetic helicity (Run~C)
and with kinetic helicity (Run~D).
These two runs are similar in that they have a much
shallower inertial range spectrum proportional to $k^{-1}$.
In addition, we also consider two cases with magnetic fields,
which are driven by an electromotive force.
The functional form is the same as for the forcing in the momentum
equation, but the amplitude factor $f_0$ tends to be much smaller
while still leading to comparable stresses.
This is a consequence of this forcing occurring underneath
the curl of the induction equation.
Run~E has null magnetic helicity and Run~F has maximum
magnetic helicity; see \Fig{pencil_hel}.
In these two cases, we find a somewhat steeper 
inertial range spectrum proportional to $k^{-5/3}$,
but for Run~F there is a clear uprise of the spectrum
at low wavenumber, which is likely a consequence
of an inverse cascade; see also \cite{Christensson:2000sp}
for hydromagnetic decay simulations showing inverse cascading.
Those cases are rather similar in that now the tensor component is
dominant; see \Fig{pencil_hel}.

As in the case of acoustic turbulence, the vector mode dominates over
the scalar mode at the largest length scales ($k<400$),
with Run~F being the exception (where the vector and scalar
modes are comparable, and both much smaller than the tensor mode).
For vortical hydrodynamic turbulence, the tensor and vector modes are comparable
in the inertial range, while for magnetic turbulence the tensor mode dominates
over this regime.
Hence, in the magnetic turbulence case, the spectrum of the
total unprojected stress tensor agrees surprisingly well with that of
the tensor mode (for all $k$ in the helical case, and at large $k$ for
the non-helical case), so the computational effort for doing such a projection
could potentially be ignored in a first approximation,
as was done in Refs.~\cite{Gogoberidze:2007an,Kahniashvili:2008er}.
This is, however, not the case for acoustic turbulence, where the
projection is essential.

It is worth noting that all spectra of \Figss{pencil_ac}{pencil_hel}
show a subinertial range proportional to $k$ or $k^{3/2}$,
which is shallower than the generic $k^2$ spectrum.
In the present case, the subinertial range is not long enough to make
strong claims, but it is useful to recall that deviations from a $k^2$
spectrum of the stress have been found when there are departures from
Gaussianity of the underlying magnetic or velocity fields \cite{Brandenburg:2019uzj}.
Although the departures from Gaussianity may not yet be very strong
in the present simulations, it is worth noting that real astrophysical
magnetic fields have much larger magnetic Reynolds numbers and may
well lead to much more significant departures.

\subsection{On the Relative GW Strength for Acoustic and Vortical Turbulence}
\label{secRelPower}

In this connection, it is interesting to note that the apparent strength
of GWs for acoustic turbulence cannot straightforwardly be explained by
the SVT decomposition.
If there was such a connection, one would have expected the
GW energy to be less for acoustic turbulence than for vortical turbulence,
because the tensor mode (i.e., the actual source for the GWs)
is weaker there.
However, this is not the case \cite{Pol:2019yex}.
We must therefore conclude that this difference between acoustic and
vortical turbulence is explained by other factors, most likely the
detailed aspects of the temporal variation of the stress. 
In fact, to drive GWs, the temporal stress frequency must be close
to the frequency of light at each spatial Fourier mode and it seems
plausible that this is more easily satisfied for acoustic turbulence.
It can be argued that only the parallel component of the perturbations in
magnetic fields contributes to the energy spectrum of the stress tensor
of $B^2$ sourcing the GWs. 
Thus the acoustic modes of turbulence are more efficient in transferring energy to the GWs than
the vortical modes, and thus GWs sourced by acoustic turbulence carry
more energy compared to those by vortical turbulence.

\begin{figure}[t!]\begin{center}
\includegraphics[width=0.8\columnwidth]{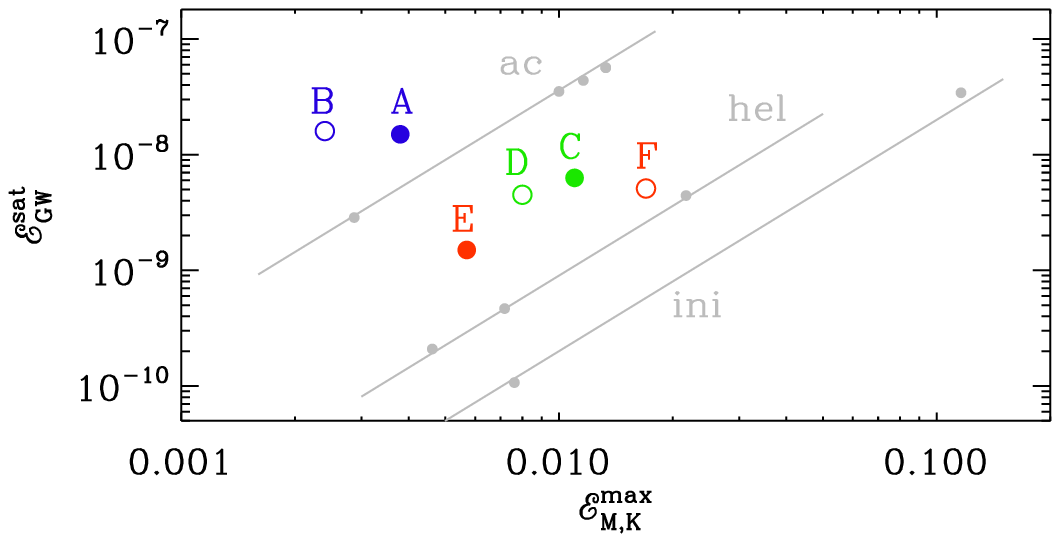}
\end{center}\caption[]{
Positions of Runs~A--F in a diagram showing
$\calEE_{\rm GW}^{\rm sat}$ versus $\calEE_{\rm M}^{\max}$,
For orientation the data points of the Ref.~\cite{Pol:2019yex}
are shown as gray symbols for acoustic (ac), helical (hel),
initialized (ini) turbulence.
}\label{EEGW_vs_EEKM}\end{figure}

In \Fig{EEGW_vs_EEKM} we show the positions of Runs~A--F in a
$\calEE_{\rm GW}^{\rm sat}$ versus $\calEE_{\rm M}^{\max}$ diagram.
For orientation, we also show the old data points from the Ref.~\cite{Pol:2019yex}.
We see that the new acoustic or irrotational turbulence simulations
are now even slightly more efficient in driving GWs: Run~B has not
only the smallest kinetic energy of all the runs, but it also
has a relatively large GW energy.

\begin{figure}[t!]\begin{center}
\includegraphics[width=0.49\columnwidth]{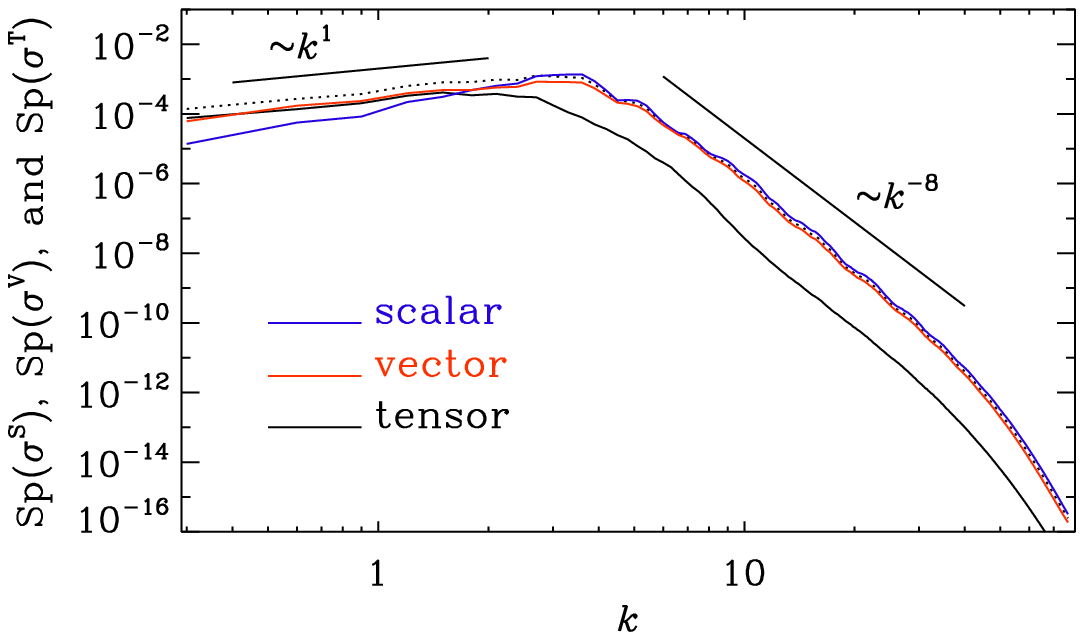}
\includegraphics[width=0.49\columnwidth]{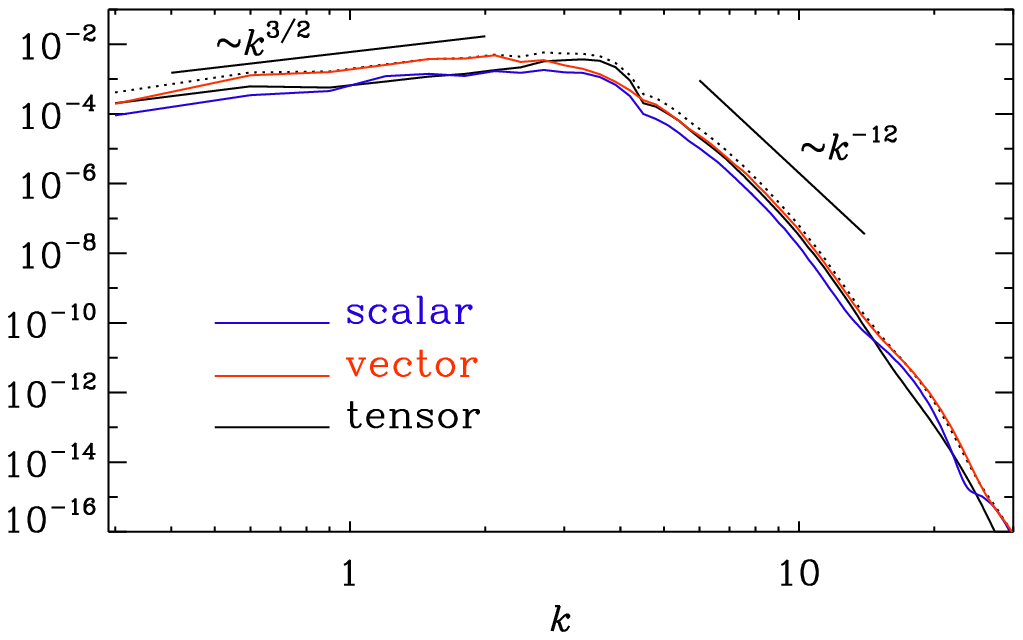}
\end{center}\caption[]{
SVT decomposition for Runs E (left) and F (right) with $\kf=2$
using irrotational and vortical forcings, respectively. 
}\label{pencil_other}\end{figure}

Larger GW energies can be obtained by decreasing $k_*$.
In \Fig{pencil_other}, we show the results for $k_*=2$ and compare
vortical and irrotational hydrodynamic forcings.
Similarly to the cases with larger $k_*$, we see, again, that
the tensor mode of acoustic turbulence is weaker (second panel of
\Fig{pencil_other}), but this is only true of the initial range,
which contributes little to the total GW energy production.
Most of the GW energy comes from wave numbers $k$ below the peak,
and there the tensor mode is, similarly to the cases with larger
$k_*$, the largest (left panel of \Fig{pencil_other} around the peak,
and right panel for low $k$) or, at least comparable to the vector mode,
which is the largest (left panel of \Fig{pencil_other} for low $k$).

\section{Relation Between Stress and Strain Spectra}
\label{secDegPol}

\subsection{GW Equation for the Radiation Dominated Epoch}

In this section, we connect the observable quantities related to GWs
to properties of the source obtained in the previous section.
GWs are considered to be small tensor perturbations over the 
FLRW metric, and the dynamics
of these GWs is given by the linearized Einstein equations.
The perturbed metric is taken to be,
\begin{equation}\label{eMetric}
ds^2=-dt_{\rm P}^2+a^2(t_{\rm P})[\delta_{ij}+h_{ij,{\rm P}}(\mathbf{x},t)]\,dx_i dx_j,
\end{equation}
where $\mathbf{x}$ are the comoving coordinates, $t_{\rm P}$ is the physical time,
$a(t_{\rm P})$ is the scale factor of the universe, and $h_{ij,{\rm P}}$ are the physical strains.
The equation governing the evolution of these GWs is given by \cite{Grishchuk:1974ny,Mukhanov:1990me},
\begin{equation}\label{eGWGeneral}
\left[\frac{\pa^2}{\pa t_{\rm P}^2}+3H(t_{\rm P})\frac{\pa}{\pa t_{\rm P}}-\frac{\nabla^2}{a^2}\right]
h_{ij,{\rm P}}(\mathbf{x},t_{\rm P})=16\pi G T^{\rm TT}_{ij,{\rm P}}(\mathbf{x},t_{\rm P}),
\end{equation}
where $G$ is Newton's constant, $H(t_{\rm P})=da/(a\,dt_{\rm P})$ is the Hubble rate,
$T^{\rm TT}_{ij,{\rm P}}$ is the \textit{physical} stress tensor, the
superscript $^{\rm TT}$ denotes the the transverse and traceless (TT) projection,
and $\nabla$ signifies derivatives with respect to the comoving coordinates $\mathbf{x}$.

The energy density in the magnetic fields and plasma velocity dilutes with the
expansion of the universe, and to scale out the effects of this expansion,
we need to express the GW equation in terms of comoving stress tensor
$T_{ij}=a^4\,T_{ij,{\rm P}}$, the scaled strain $h_{ij}=a\,h_{ij,{\rm P}}$, and the
conformal time $t$ defined as  $dt_{\rm P}= a\,dt$.
Also in the radiation dominated epoch, the scale factor is linear in conformal
time, and the damping term in the above equation vanishes.
Eq.~\eqref{eGWGeneral} can then be expressed as
\begin{equation}\label{eGWRadComov}
h''_{ij}-\nabla^2 h_{ij}=\frac{16\pi G}{a}T^{\rm TT}_{ij}(\mathbf{x},t),
\end{equation}
where the primes denote derivatives with respect to the conformal time $t$.
The analytic solution of the GW equation is discussed in several references;
see, for example, Refs.~\cite{Gogoberidze:2007an, Kahniashvili:2005qi}.
The full numerical solution for a turbulent source is also discussed in
Refs.~\cite{Pol:2018pao, Pol:2019yex}.

The four-dimensional power spectrum $H_{ijlm}(\mathbf{k},t,\tau)$ of the energy
density tensor is defined as
\begin{equation}\label{eHijlmKandT}
\frac{1}{w^2}\left<\tilde{T}^*_{ij}(\mathbf{k},t)\tilde{T}_{lm}(\mathbf{k}',t+\tau)\right>=(2\pi)^3\delta^3(\mathbf{k}-\mathbf{k}')H_{ijlm}(\mathbf{k},t,\tau),
\end{equation}
where $\tilde{T}_{ij}(\mathbf{k},t)$ is the Fourier transform of
$T_{ij}(\mathbf{x},t)$, we have used the reality condition
$\tilde{T}_{ij}(-\mathbf{k},t)=\tilde{T}^*_{ij}(\mathbf{k},t)$.
As mentioned previously, $w\equiv\rho+p$ is the enthalpy density of the primordial plasma,
where $\rho$ and $p$ being the energy density and the pressure of the primordial
plasma, and all quantities are comoving.

In the standard framework there are two degrees of freedom for $h_{ij}$
(a $3 \times 3$ tensor has 9 components, the symmetry condition $h_{ij}=h_{ji}$
leaves 6 degrees of freedom, the transverse nature of GWs, i.e., $k_i h^{ij} =0$,
reduces that to 3, while the traceless condition $h_{ii}=0$ further reduces the
total number of the degrees of freedom to 2) that correspond to two possible states
of GWs polarization.
In the case of a parity even source, these two states are equal to each other,
while a parity odd sources generates circularly polarized GWs, where there
is an excess of one state over the other.
The circular polarization of GWs can be understood intuitively by working in the
\textit{circular polarization basis}, which can be constructed from the unit
vector $\hat{{\bf k}}$, and two more unit vectors $\hat{{\bf e}}^1$ and
$\hat{{\bf e}}^2$ which are perpendicular to $\hat{{\bf k}}$ and to each other.
The basis is defined as \cite{Caprini:2003vc}
\begin{equation}\label{eCircPolBas}
\hat{e}^{R,L}_{ij}(\hat{\mathbf{k}})=-\frac{1}{2}\big(\hat{e}_1\pm i\hat{e}_2\big)_i\big(\hat{e}_1\pm i\hat{e}_2\big)_j.
\end{equation}
It is easy to check that $(e_{ij}^L)^*=e_{ij}^R$, and vice versa; and also that 
\begin{eqnarray}\label{eCircPolBasOrtho}
	\hat{e}^{L}_{ij}(\hat{\mathbf{k}})\hat{e}^{R}_{ij}(\hat{\mathbf{k}})&=1,\\
	\hat{e}^{L}_{ij}(\hat{\mathbf{k}})\hat{e}^{L}_{ij}(\hat{\mathbf{k}})&=
	\hat{e}^{R}_{ij}(\hat{\mathbf{k}})\hat{e}^{R}_{ij}(\hat{\mathbf{k}})=
	\hat{\mathbf{k}}_i\hat{e}^{R,L}_{ij}=\delta_{ij}\hat{e}^{R,L}_{ij}=0. 
\end{eqnarray}
The Fourier space stress tensor can now be decomposed into the left and right
handed components,
\begin{equation}\label{eCircPolBasDec}
\tilde{T}_{ij}(\mathbf{k},t)=\tilde{T}_L(\mathbf{k},t) \,e^L_{ij}(\hat{{\bf k}}) + \tilde{T}_R(\mathbf{k},t)\,e^R_{ij}(\hat{{\bf k}}).
\end{equation}
We can also construct symmetric and antisymmetric projectors \cite{Caprini:2003vc},
\begin{eqnarray}\label{eProjOps}
\mathcal{S}_{ijlm}(\hat{\mathbf{k}})&\equiv \frac{1}{2}\left[e^L_{ij} e^R_{lm}+e^R_{ij} e^L_{lm}\right],\\
\mathcal{A}_{ijlm}(\hat{\mathbf{k}})&\equiv \frac{1}{2i}\left[e^R_{ij} e^L_{lm}-e^L_{ij} e^R_{lm}\right].
\end{eqnarray} 
These projection operators satisfy the properties that $\mathcal{S}_{ijlm}(\hat{\mathbf{k}})\mathcal{S}_{ijlm}(\hat{\mathbf{k}})
=\mathcal{A}_{ijlm}(\hat{\mathbf{k}})\mathcal{A}_{ijlm}(\hat{\mathbf{k}})=2$, and
$\mathcal{S}_{ijlm}(\hat{\mathbf{k}})\mathcal{A}_{ijlm}(\hat{\mathbf{k}})=0$,
and pick out respectively the symmetric and antisymmetric components of a
fourth-order tensor.

The degree of polarization of the \textit{source} can then be expressed as
\begin{equation}\label{eDegPolSource}
\mathcal{P}_T(k,t)=\frac{T^\mathcal{A}(k,t)}{T^\mathcal{S}(k,t)},
\end{equation}
where
\begin{equation}\label{eDegPolSourceInts}
T^\alpha(k,t)=\frac{1}{2w^2}\int_{\Omega_k} d\Omega_k \, k^2\,\int d^3\mathbf{k}'\alpha_{ijlm}(\hat{\mathbf{k}})\left<\tilde{T}^*_{ij}(\mathbf{k},t)\tilde{T}_{lm}(\mathbf{k}',t)\right>,
\end{equation}
with $\alpha$ denoting either $\mathcal{S}$ or $\mathcal{A}$.

In terms of the polarization components of Eq.~\eqref{eCircPolBasDec}, we can write
\begin{eqnarray}\label{eDegPolSourceCirc}
T^\mathcal{S}(k)&=\frac{1}{2w^2}\int_{\Omega_k} d\Omega_k \, k^2\,\int d^3\mathbf{k}'\left<\tilde{T}^*_R(\mathbf{k})\tilde{T}_R(\mathbf{k}')+\tilde{T}^*_L(\mathbf{k})\tilde{T}_L(\mathbf{k}')\right>,\\
T^\mathcal{A}(k)&=\frac{1}{2w^2}\int_{\Omega_k} d\Omega_k \, k^2\,\int d^3\mathbf{k}'\left<\tilde{T}^*_R(\mathbf{k})\tilde{T}_R(\mathbf{k}')-\tilde{T}^*_L(\mathbf{k})\tilde{T}_L(\mathbf{k}')\right>,
\end{eqnarray}
where the explicit $t$-dependence of the quantities in Eq.~\eqref{eDegPolSourceCirc}
is dropped for brevity.
$\mathcal{P}_T(k,t)$ thus captures the relative inequality of power in the left
and right handed components of the stress tensor.
One can alternatively work in the \textit{linear polarization basis}, where the physical
meaning of the polarization degree of the source is not so straightforward.
The definition of the basis, as well as the expressions analogous to those above
are presented in \ref{AppLinPolBas}.
For more details about the linear and circular polarization bases, and the
decomposition of the four-dimensional power spectrum into its symmetric
and antisymmetric parts, see Ref.~\cite{Caprini:2003vc}, where the stress
tensor is sourced by magnetic fields. 
Notably, the analytical estimates for the polarization degree differ
significantly from the numerical simulations due to the simplified 
treatment of the time-decorrelation functions and the neglect of different 
scaling laws for the energy density and helicity decays scalings; see
Ref.~\cite{Kahniashvili:2020jgm} for more discussion. 

Since GWs are sourced by the TT projected stress tensor, i.e., the tensor
component $\bar{\lambda}_{ij}$ of the stress tensor [cf.\ Eq.~\eqref{eSVTMain}],
we built the four-dimensional power spectrum $H_{ijlm}(\mathbf{k},t,\tau)$
from this TT component and decomposed it into parity odd and even parts.
It is of interest to see whether it is possible to ascribe a degree of polarization
analogous to that in Eq. \eqref{eDegPolSource} to the scalar and vector components
of the stress tensor.
We defer a discussion of that to \ref{AppPolSV}

\subsection{Shell-integrated Spectra}

In a manner analogous to Eq.~\eqref{eHijlmKandT}, we can form a four-dimensional
power spectrum $Q_{ijlm}(\mathbf{k},t,\tau)$ from the time derivative of the strains
in Fourier space
\begin{equation}\label{eQijlmKandT}
\frac{1}{32\pi G}\left<\dot{\tilde{h}}^*_{ij,{\rm P}}(\mathbf{k},t)\dot{\tilde{h}}_{lm,{\rm P}}(\mathbf{k}',t+\tau)\right>=(2\pi)^3\delta^3(\mathbf{k}-\mathbf{k}')Q_{ijlm}(\mathbf{k},t,\tau).
\end{equation}
For more details on the fourth order correlation functions and their Fourier
transforms, see Ref.~\cite{monin-yaglom}.

The polarization degree of GWs is written as
\begin{equation}\label{eDegPolGW}
\mathcal{P}_{GW}(k,t)=\frac{H_{GW}(k,t)}{E_{GW}(k,t)},
\end{equation}
where $E_{GW}(k,t)$ and $H_{GW}(k,t)$ are respectively the spectral GW
energy density and the spectral GW helical energy density, obtained
respectively by the application of the projection operators of Eq.~\eqref{eProjOps}
on the LHS of Eq.~\eqref{eQijlmKandT}, in a way similar to Eq.~\eqref{eDegPolSourceInts}.

In the circular polarization basis, we get
(dropping the explicit $t$-dependence)
\begin{eqnarray}\label{eEandHGwSpecsCirc}
E_{GW}(k)&=\frac{1}{(2\pi)^6}\frac{1}{32\pi G}\int_{\Omega_k} d\Omega_k \, k^2\,\int d^3\mathbf{k}'\nonumber\\
&\times\left<\dot{\tilde{h}}^*_R(\mathbf{k})\dot{\tilde{h}}_R(\mathbf{k}')+\dot{\tilde{h}}^*_L(\mathbf{k})\dot{\tilde{h}}_L(\mathbf{k}')\right>,\\
H_{GW}(k)&=\frac{1}{(2\pi)^6}\frac{1}{32\pi G}\int_{\Omega_k} d\Omega_k \, k^2\,\int d^3\mathbf{k}'\nonumber\\
&\times\left<\dot{\tilde{h}}^*_R(\mathbf{k})\dot{\tilde{h}}_R(\mathbf{k}')-\dot{\tilde{h}}^*_L(\mathbf{k})\dot{\tilde{h}}_L(\mathbf{k}')\right>,
\end{eqnarray}
where $\dot{\tilde{h}}_{ij,{\rm P}}(\mathbf{k},t)$ is decomposed in the circular
polarization basis,
\begin{equation}\label{eCircPolBasDecHdot}
\dot{\tilde{h}}_{ij}(\mathbf{k},t)=\dot{\tilde{h}}_L(\mathbf{k},t) \,e^L_{ij}(\hat{{\bf k}}) + \dot{\tilde{h}}_R(\mathbf{k},t)\,e^R_{ij}(\hat{{\bf k}}).
\end{equation}
As expected, $\mathcal{P}_{GW}(k,t)$ captures the relative inequality of power
in the left and right handed components of the GW perturbations.

\begin{figure}[t!]\begin{center}
\includegraphics[width=\columnwidth]{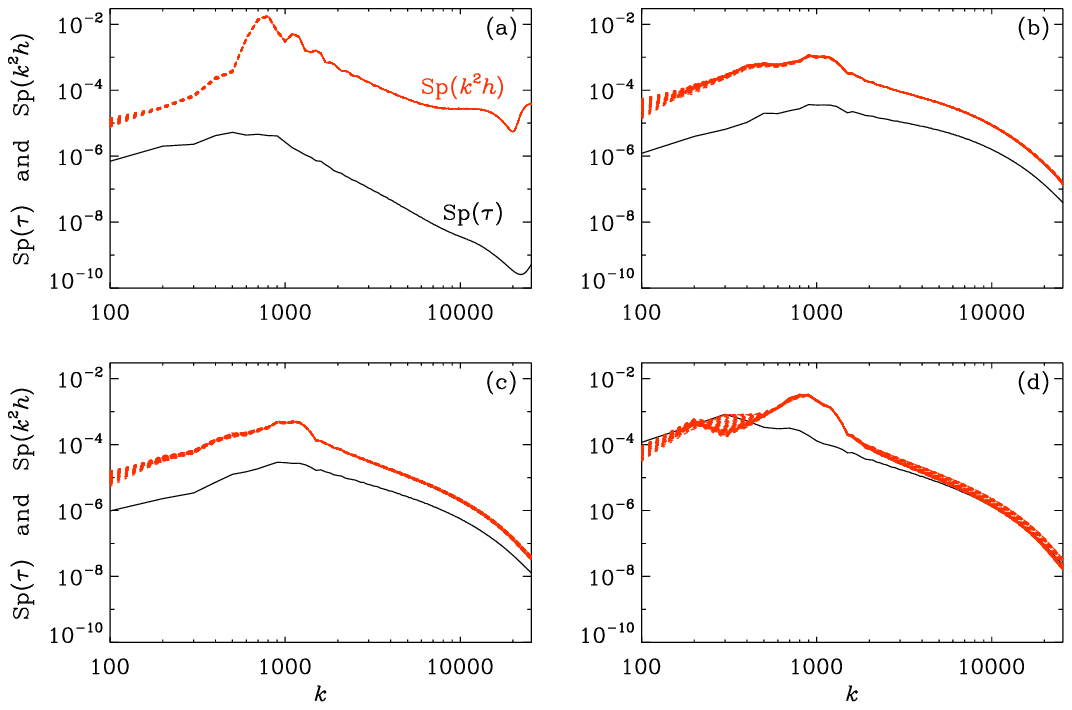}
\end{center}\caption[]{
Spectra of stress (solid line at maximum) and strain
(dotted lines for different times after the maximum)
for Runs~B, C, E, and F.
}\label{pstress_strain4}\end{figure}

\subsection{Numerical Results for the Strain and Stress Spectra}

We now solve for the resulting GWs numerically using the approach
described in Refs.~\cite{Pol:2018pao,Pol:2019yex}.
Here we focus on a comparison of the results with irrotational plane
wave (Run~B) and vortical plane wave forcings without (Run~C) and with
(Run~D) magnetic field, and with helicity (Run~F).
In each of the cases, we show the spectra of both the TT-projected stress
of the magnetic (kinetic) fields $\bar{\tau}_{ij}$ ($\bar{\sigma}_{ij}$),
and the strain $h$ scaled with a $k^2$ factor.
Here we denote the latter simply as $\Sp(k^2h)$, but mean by this all
of the components of the strain tensor.
Likewise, we denote the stress spectrum by $\Sp(\tau)$, and mean by
this the TT-projected stress tensor with the $16\pi G$ factor included,
so that it would be equal to $\Sp(k^2h)$, if the second conformal time
derivative of the strain in \Eq{eGWRadComov} was absent.
The result is shown in \Fig{pstress_strain4} for Runs~B, C, E, and F.

We see from \Figp{pstress_strain4}{d} that in the magnetic case with
helicity, the spectra of $\Sp(k^2h)$ and $\Sp(\tau)$ are in almost
perfect agreement with each other at all wave numbers, except near $2\kf$.
The situation changes when there is no helicity and $\Sp(k^2h)$ is now
slightly above $\Sp(\tau)$; see \Figp{pstress_strain4}{c}.
The difference increases further when we compare with purely hydrodynamic
cases (Run~C); see \Figp{pstress_strain4}{b}.
Particularly striking is a difference for irrotational turbulence, where
we see significant differences between the two spectra at all wave numbers
(Run~B); see \Figp{pstress_strain4}{a}.

\begin{figure}[t!]\begin{center}
\includegraphics[width=\columnwidth]{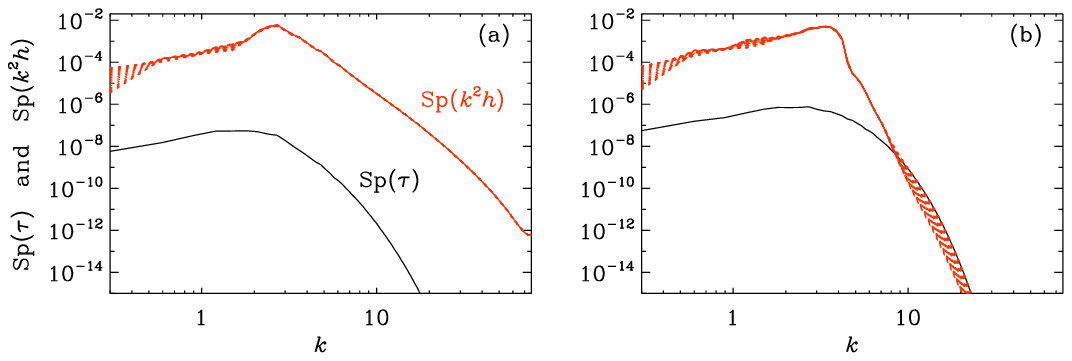}
\end{center}\caption[]{
Spectra of stress (solid line at maximum) and strain
(dotted lines for different times after the maximum)
for Runs~G and H.
}\label{pstress_strain}\end{figure}

Our results suggest that, although there is a correspondence between
$\Sp(k^2h)$ and $\Sp(\tau)$, there can also be a significant departure
and even a shift between them; see \Fig{pstress_strain}.
Thus, even though the TT-projection of the stress has resulted in a
significant decrease in the projected stress (see \Fig{pencil_ac}),
the stress in the irrotational hydrodynamic cases have now other
qualities that make it more efficient for driving GWs.
As already discussed above, this could be caused by larger power at
larger temporal frequencies.
This is supported by our findings that in the cases with the
smallest efficiency (Run~D), the two spectra are almost equal,
i.e., $\Sp(\tau)\approx\Sp(k^2h)$.

\subsection{Polarization Spectra of Stress and Strain}

It is common to define the degree of polarization in terms of the
strains themselves, as opposed to our definition in terms of the time derivatives
of the strains (see, e.g., Eq.~(6) of Ref.~\cite{Kahniashvili:2005qi}).
It can be argued that these definitions are equivalent for GWs sourced by
stationary turbulence, where one can apply the \textit{aeroacoustic approximation}.
For stationary turbulence, one can take the time derivatives out of the time
averaging operation on the four-point correlation function of the stress tensor
(see Eq.~(12) of Ref.~\cite{Gogoberidze:2007an}).
The same reasoning applies to the four-point correlation function of the GW
strain, since the strains are expected to have the same statistical properties
as the turbulent source.
This means
\begin{eqnarray}\label{eAeroHij}
&\left<\dot{h}^{\rm TT}_{ij,{\rm P}}(\mathbf{x},t)  \dot{h}^{\rm TT}_{lm,{\rm P}}(\mathbf{x}+\bm{\xi},t+\tau)\right>_t \nonumber\\
&=-\pa_t^2\left<h^{\rm TT}_{ij,{\rm P}}(\mathbf{x},t)  h^{\rm TT}_{lm,{\rm P}}(\mathbf{x}+\bm{\xi},t+\tau)\right>_t,
\end{eqnarray}
where the subscript $t$ is used to denote that the angle brackets now also
include time averaging, in addition to the ensemble averaging.
In frequency space, the time derivatives become factors of frequency, and cancel
out when calculating the degree of polarization (see Eq.~\eqref{eDegPolGW}).

\begin{figure}[t!]\begin{center}
\includegraphics[width=\textwidth]{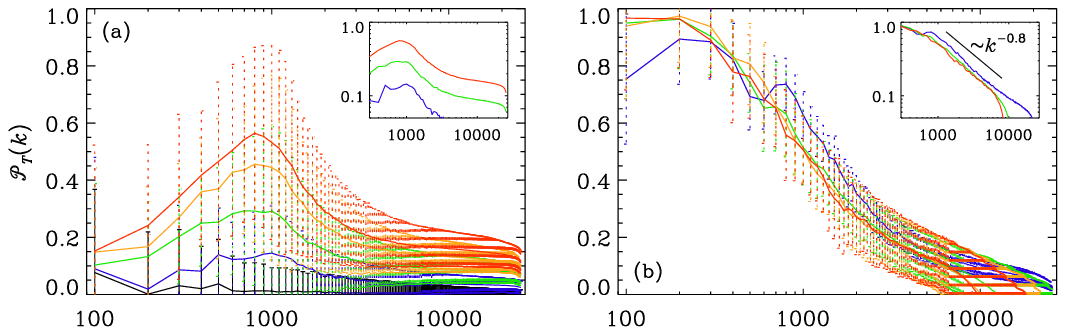} 
\includegraphics[width=\textwidth]{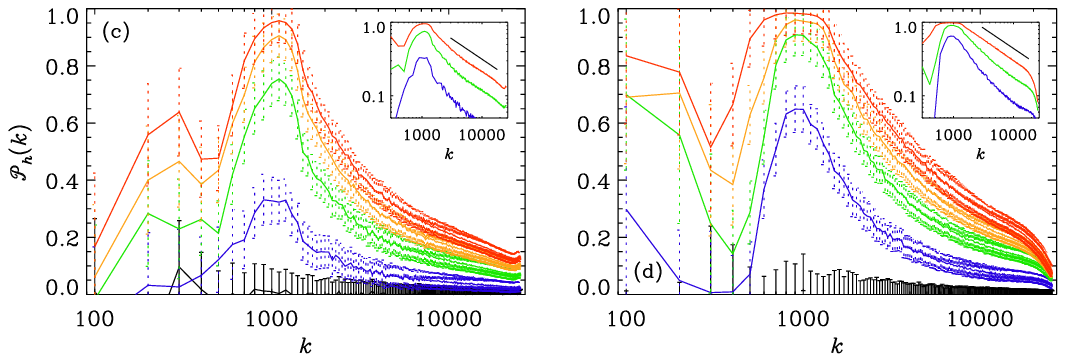}
\end{center}\caption[]{
Degree of polarization of $T^{\rm TT}$ for (a) Runs~K0, K01, K03, K05, and K1 of 
Ref.~\cite{Kahniashvili:2020jgm}, and (b) their Runs~M01, M03, M05, and M1,
as well as the degree of polarization of $h$ (c) for Runs~K0, K01, K03, K05, and K1,
and (d) for Runs~M01, M03, M05, and M1.
The inserts show the same, but in a double-logarithmic representation.
For (b), the straight line has a slope of $-0.8$, while for
(c) and (d) the two straight lines have a slope of $-0.55$.
}\label{pstre_comp_hor}\end{figure}

Let us now discuss the degree of polarization of the source, 
${\cal P}_T(k)$, and of the strain. ${\cal P}_h(k)$, from numerical simulations.
In \Fig{pstre_comp_hor}(a) and (b) we show ${\cal P}_T(k)$ for the source for
the runs of Ref.~\cite{Kahniashvili:2020jgm} for kinetic and magnetic driving.
The run names for the kinetically driven cases are K0, K01, K03, K05, and K1
for $\sigma=0$, $0.1$, $0.3$, $0.5$, and $1$, respectively. 
The magnetically driven runs have analogous names M0, M01, M03, M05, and M1.

Remarkably, in the kinetically driven case, the polarization of the
source never reaches 100\%, and yet, the polarization of $h$ does reach
nearly 100\%. 
This was already found in Ref.~\cite{Kahniashvili:2020jgm}, and their
results are convenient comparison also reproduced in
\Fig{pstre_comp_hor}(c) and (d). 
Furthermore, in the magnetically driven case, the source reaches nearly 
100\%, but not at the same wave number where the GWs become
100\% polarized.
Instead, maximum polarization happens here at the smallest scale.

\section{Conclusions}
\label{secConcl}

We have computed the SVT decomposition for six different flows:
kinetically and magnetically driven flows, with and without
helicity, as well as two types of acoustically driven flows. 
In addition, we have considered two cases with significantly smaller
forcing wave numbers. 
All our flows had a turbulent inertial range spectrum, as well as a
subinertial range spectrum.
However, significant differences can be seen in the relative
strengths of scalar, vector, and tensor modes of the turbulent source.
The largest contribution to the tensor mode comes from the flows
in Runs~E and F.
Both are magnetically driven, although only one of them had helicity.
In the presence of magnetic helicity (Run~F), an inverse cascade emerges,
which leads to an additional contribution to the tensor mode at small
wave numbers. 
The scalar and vector modes of the turbulence stress
are small and of comparable strengths
in the helically driven case.
Without helicity, however, the vector mode exceeds the scalar mode at
small and intermediate length scales.
For helically driven flows, there is no significant inverse cascade,
and the tensor mode is no longer so dominant.
Instead, tensor and vector modes are now of comparable strength.
At smaller wave numbers, the vector mode does even become dominant.

For acoustically or irrotationally driven flows, the situation is
drastically different in that the tensor mode is now comparatively weak,
while scalar and vector modes are dominant and of similar strengths.
Furthermore, there is very little difference between two fundamentally different
realizations of acoustic turbulence: plane wave and spherical expansion
wave forcings.
Both show a very strong ``bottleneck'' effect, i.e., a shallower spectrum
at high wave numbers, which was already found in Ref.~\cite{Mee:2006mq}.
Bottleneck effects are known to exist also for vortical turbulence
\cite{Falk94}, but they are not as strong \cite{Dobler:2003xh}.
The planar wave forcing was thought to be more similar to the vortical
plane wave forcing, but this is not the case.
Even though the acoustically driven flows have comparatively weak tensor
modes, they are most effective in driving GWs. 
The reason for this higher efficiency of acoustic flows in generating 
GWs is not very well understood; but it might be 
interesting to note that at least in a gravitationally stratified fluid, 
maximum acoustic wave power output was found for flows 
with significant longitudinal modes compared to transverse ones \cite{Stein:1967}. 
The planar wave-driven acoustic flows are even slightly more effective
(Run~B, $q=32$), than spherical expansion wave-driven flows (Run~A, $q=19$).

Given that the presence of helicity had no significant effect on the
results of the SVT decomposition, it was of interest to look
for differences in the degree of polarization of the source and
the resulting GWs.
The difference is indeed immense, but it results almost entirely from the
inverse cascade, which is only possible in the magnetically driven case.
In these cases, the degree of polarization of the source
is virtually independent of the fractional helicity.
This is probably because, even for weak magnetic helicity, inverse
cascading is possible.
At the driving scale, however, the degree of polarization is never
close to 100\% -- neither in the magnetically driven cases, nor in the
kinetically driven ones.
In the latter, however, fractional helicity of the velocity does lead
to a smaller fractional helicity of the stress.
By contrast, in the magnetically driven cases, all spectra of the
polarization degree are similar and they seem to be controlled by some
kind of forward cascade from the largest scale downward to all smaller
scales.
This includes even the driving scale of the system, whose degree of
polarization seems to be controlled by the stress on the largest scale
of the domain.

Clearly, spectral power on the scale of the domain must be interpreted
with care, because the intention of periodic box simulations is to
represent an infinite and unbounded domain.
The results will therefore be different if one considered a larger domain.
It is conceivable, then, that inverse cascading also works in an
infinitely extend domain, but, again, only on the scale of the domain
will the polarization degree be close to maximum.

The work presented here will be important for identifying the nature of the 
turbulent motions in the early universe once a GW signal from turbulent 
sources will be detected. 
More specifically, it can help to disentangle the turbulent conditions
in the early universe regarding the dependence on 
irrotational (longitudinal) and vortical (solenoidal) flows.

\vspace{2mm}
Data availability---The source code used for the
simulations of this study, the {\sc Pencil Code},
is freely available \cite{PC}.
The simulation setups and the corresponding data
are freely available from Refs.~\cite{DATA,DATA2}.

\section*{Acknowledgments}

We thank Arthur Kosowsky and A.~G.~Tevzadze for useful discussions.
We also acknowledge the two anonymous referees for useful suggestions.
Support through the Swedish Research Council, grant 2019-04234,
and Shota Rustaveli GNSF  (grants FR/18-1462)
are gratefully acknowledged.
We acknowledge the allocation of computing resources provided by the
Swedish National Allocations Committee at the Center for Parallel
Computers at the Royal Institute of Technology in Stockholm.

\section*{ORCID iDs}
Axel Brandenburg: \url{https://orcid.org/0000-0002-7304-021X} \\
Grigol Gogoberidze: \url{https://orcid.org/0000-0002-0506-333X} \\
Tina Kahniashvili: \url{https://orcid.org/0000-0003-0217-9852} \\
Sayan Mandal: \url{https://orcid.org/0000-0003-3129-5799} \\
Alberto Roper Pol: \url{https://orcid.org/0000-0003-4979-4430} \\

\appendix
\section{Helmholtz Theorem and the real space SVT decomposition}
\label{AppHelm}

The scalar-vector-tensor (SVT) decomposition is the most generalized way to decompose
linearized perturbations of a symmetric rank-2 tensor based on how these components transform
under rotations.
In this formulation, the evolution equations for the scalar, vector, and tensor components
of the metric perturbations are decoupled from each other at the linear level.
It is a more generalized version of the Helmholtz decomposition by which a vector is
decomposed into divergence-free and curl-free parts.

In Sec.~\ref{secSVT}, we have provided the expressions for the SVT decomposition of a
general symmetric rank-2 tensor in momentum space.
In this Appendix, we show the corresponding decomposition in real space.

The Helmholtz theorem states that any 3-vector $b_i$ can be decomposed into
a scalar part $b$ and a vector part $\bar{b}_i$,
\begin{equation}\label{eHelmDec}
b_i=\nabla_i b+\bar{b}_i.
\end{equation}
The first term on the RHS is curl-free and the second term is divergence-free, $\nabla_i \bar{b}_i=0$.
To extract this latter part, we need to solve for $b$; we get the following condition on $b$,
\begin{equation}\label{eExtractDivLessVect}
\nabla^2 b=\nabla_i b_i(\mathbf{x}).
\end{equation}
This corresponds to a Poisson's equation, whose general solution in three dimensions is
\begin{equation}\label{eSolForB}
b=-\int dV' \,\frac{\nabla'_i b_i (\mathbf{x}')}{4\pi|\mathbf{x}-\mathbf{x}'|},
\end{equation}
where the prime on the $\nabla$ indicates that the derivative is with respect to the
components of $\mathbf{x}'$; we thus get,
\begin{equation}\label{eDivLessVec}
\bar{b}_i=b_i+\nabla_i\int dV' \,\frac{\nabla'_j b_j (\mathbf{x}')}{4\pi|\mathbf{x}-\mathbf{x}'|}.
\end{equation}

Analogous to the case for vectors, we have SVT decomposition for tensors
-- any symmetric rank-2 tensor $\lambda_{ij}(\mathbf{x})$ can be decomposed
as in Eq.~\eqref{eSVTMain}.
Taking the trace of Eq.~\eqref{eSVTMain}, we get,

\begin{equation}\label{eLambSVT-Trace}
L=\frac{1}{d}\lambda_{ii},
\end{equation}
which allows us to rewrite Eq.~\eqref{eSVTMain}, using the condensed notation $\lambda_{\langle ij\rangle}\equiv\lambda_{ij}-\delta_{ij}\lambda_{ss}/d$,
\begin{equation}\label{eLambSVT-2}
\lambda_{\langle ij\rangle}=\nabla_{\langle i}\nabla_{j\rangle}\lambda+\nabla_{(i}\bar{\lambda}_{j)}+\bar{\lambda}_{ij}.
\end{equation}
Acting on this with $\nabla_i$ and then with $\nabla_j$, we get,
\begin{equation}\label{eDelSqDelSqTau}
\nabla_i\nabla_j\lambda_{\langle ij\rangle}=\frac{d-1}{d}\nabla^2\big(\nabla^2\lambda\big),
\label{Poisson}
\end{equation}
and thus, we can solve for $\lambda$, by solving the Poisson's
Eq.~\eqref{Poisson} first for $\nabla^2 \lambda$, and then for $\lambda$
in $d$ dimensions\footnote[1]{For notes on deriving the Green's function
of $\nabla^2$ in $d$ dimensions see, for example, the discussion in
\url{https://tinyurl.com/knnm7bbc}.},
\begin{equation}\label{eLamb}
\lambda=\frac{d}{d-1}\frac{\Gamma^2\left({\frac{d}{2}}-1\right)}{16\pi^d}\int dV' \int dV''\frac{\nabla''_i \nabla''_j \lambda_{\langle ij\rangle}(\mathbf{x}'')}{|\mathbf{x}-\mathbf{x}'|^{d-2}|\mathbf{x}'-\mathbf{x}''|^{d-2}}.
\vspace*{1ex}
\end{equation}
To extract $\bar{\lambda}_i$, we insert this value of $\lambda$ into Eq.~\eqref{eSVTMain} and act on it with $\nabla_j$,
\begin{equation}\label{eDelILamb2}
\nabla^2\bar{\lambda}_i=2\nabla_j\lambda_{\langle ij\rangle}-\frac{2(d-1)}{d}\nabla_i(\nabla^2\lambda\big),
\end{equation}
where $\lambda$ is understood to be known from Eq.~\eqref{eLamb}. Solving this,
\begin{equation}\label{eBarTauI}
\vspace*{1ex}
\bar{\lambda}_i=-\frac{\Gamma\left({\frac{d}{2}}-1\right)}{2\pi^{\frac{d}{2}}}\int dV'\frac{\nabla'_j\lambda_{\langle ij\rangle}(\mathbf{x}')-(d-1)\nabla'_i\nabla'^2\lambda(\mathbf{x}')/d}{|\mathbf{x}-\mathbf{x}'|^{d-2}}.
\end{equation}
Finally, the traceless, divergenceless part of the energy-momentum tensor is, 
\begin{eqnarray}\label{eTTEnMom}
\bar{\lambda}_{ij}&=\lambda_{\langle ij\rangle}-\frac{d}{d-1}\frac{\Gamma^2\left({\frac{d}{2}}-1\right)}{16\pi^d}\nabla_{\langle i}\nabla_{j\rangle}\int dV' \int dV'' \,\frac{\nabla''_m \nabla''_n \lambda_{\langle mn\rangle}(\mathbf{x}'')}{|\mathbf{x}-\mathbf{x}'|^{d-2}|\mathbf{x}'-\mathbf{x}''|^{d-2}} \nonumber\\
&+\frac{\Gamma\left({\frac{d}{2}}-1\right)}{4\pi^{\frac{d}{2}}}\int dV' \nonumber \\ 
&\cdot\Bigg[\nabla_j\frac{\nabla'_m\lambda_{\langle im\rangle}(\mathbf{x}')+\frac{\Gamma^2\left({\frac{d}{2}}-1\right)}{16\pi^d}\nabla'_i\nabla'^2\ds{\int dV'' \int dV''' \,\frac{\nabla'''_m \nabla'''_n \lambda_{\langle mn \rangle}(\mathbf{x}''')}{|\mathbf{x}'-\mathbf{x}''|^{d-2}|
\mathbf{x}''-\mathbf{x}'''|^{d-2}}}}{|\mathbf{x}-\mathbf{x}'|^{d-2}} \nonumber\\
&+ (i\longleftrightarrow j)\Bigg].\end{eqnarray}

\section{An example}
\label{Examples}

It is instructive to consider an example for the formalism of
Sec.~\ref{SVTDecompMagneticStress}.
For a Beltrami field, $\mathbf{B}=(0,\sin kx,\cos kx)$, we have
\begin{equation}\label{ExampleEqn1}
\tau_{ij}={\textstyle\frac{1}{2}}
\pmatrix{
-1 & 0 & 0 \cr
0 & -\cos2kx & \sin2kx \cr
0 &  \sin2kx & \cos2kx 
}.
\end{equation}
One can easily verify that
\begin{equation}\label{ExampleEqn2}
\bar{\tau}_{ij}={\textstyle\frac{1}{2}}
\pmatrix{
0 & 0 & 0 \cr
0 & -\cos2kx & \sin2kx \cr
0 &  \sin2kx & \cos2kx }
\end{equation}
is traceless and transverse.
Using $\tau=(y^2+z^2)/4$, we have
$\nabla_{\langle i}\nabla_{j\rangle}\tau={\rm diag}(-2,1,1)/6$.
Next, using $T=-1/6$, we see that $T\delta_{ij}
+\nabla_{\langle i}\nabla_{j\rangle}\tau={\rm diag}(-1/2,0,0)$.
Thus, following \Eq{eTauSVT}, we can write $\tau_{ij}$ as $T\delta_{ij}
+\nabla_{\langle i}\nabla_{j\rangle}\tau+\nabla_{(i}\bar{\tau}_{j)}
+\bar{\tau}_{ij}$ with $\bar{\tau}_j=0$.
The Beltrami field has therefore only a spatially periodic tensor mode,
a spatially (unbounded) nonperiodic scalar mode, and no vector mode.

\section{The SVT decomposition of the fluid stress tensor}
\label{AppSVTFluid}
In this section, we provide the expressions for the scalar, vector, and tensor
components of the plasma velocity stress tensor of Eq.~\eqref{eStressFirstVel}.
As before, one gets the trace-term to be
\begin{equation}\label{eSVTVelScalar-1}
S=-p+\frac{1}{3}w \gamma^2 u^2,
\end{equation}
and we can then solve for $\sigma$,
\begin{equation}\label{eSig}
\sigma=\frac{3}{32\pi^2}\int dV' \int dV'' \,\frac{\nabla''_i \nabla''_j \left\{\left(p+\rho\right)(\mathbf{x}'')
\gamma^2 (\mathbf{x}'') u_{\langle i}u_{j\rangle}(\mathbf{x}'') \right\}}{|\mathbf{x}-\mathbf{x}'||\mathbf{x}'-\mathbf{x}''|}.
\end{equation}
The vector part is also extracted in a straightforward manner,
\begin{equation}\label{eBarSigmaI}
\bar{\sigma}_i=-\frac{1}{2\pi}\int  dV'\frac{\nabla'_j\left\{\left(p+\rho\right)(\mathbf{x}')
\gamma^2 (\mathbf{x}') u_{\langle i}u_{j\rangle}(\mathbf{x}')\right\}-2\nabla'_i\nabla'^2\sigma(\mathbf{x}')/3}{|\mathbf{x}-\mathbf{x}'|}.
\end{equation}
The tensor part finally comes out by subtracting the scalar and vector parts from the full tensor,
\begin{equation}\label{eBarSigmaIJ}
\bar{\sigma}_{ij}=w \gamma^2 u_{\langle i}u_{j\rangle}-\nabla_{\langle i}\nabla_{j\rangle}\sigma-\nabla_{(i}\bar{\sigma}_{j)}.
\end{equation}

\section{The linear polarization basis}
\label{AppLinPolBas}

In a manner similar to the circular polarization basis, one can also define the
\textbf{linear polarization basis} \cite{Caprini:2003vc},
\begin{equation}\label{eLinPolBas}
e^+_{ij}(\hat{{\bf k}})=e^1_i e^1_j-e^2_i e^2_j,\quad e^{\times}_{ij}(\hat{{\bf k}})=e^1_i e^2_j+e^2_i e^1_j,\\
\end{equation}
which is orthogonal, but not normalized, and has the properties
\begin{equation}\label{eLinPolBasOrth}
e^+_{ij}(\hat{{\bf k}})e^+_{ij}(\hat{{\bf k}})=e^{\times}_{ij}(\hat{{\bf k}})e^{\times}_{ij}(\hat{{\bf k}})=2,\quad e^+_{ij}(\hat{{\bf k}})e^{\times}_{ij}(\hat{{\bf k}})=0.
\end{equation}
In addition, the fact that $\hat{{\bf e}}^1$, $\hat{{\bf e}}^2$, and $\hat{{\bf k}}$ form a right-handed orthonormal basis leads to the relations
\begin{equation}\label{eLinPolBasOrthInter}
-\epsilon_{ipq}\hat{k}_p e^{\times}_{qj}(\hat{{\bf k}})=e^+_{ij}(\hat{{\bf k}}),\quad \epsilon_{ipq}\hat{k}_p e^+_{qj}(\hat{{\bf k}})=e^{\times}_{ij}(\hat{{\bf k}}).
\end{equation}
As in Eq.~\eqref{eCircPolBasDec}, we can decompose the stress tensor into
$+$ and $\times$ polarization modes,
\begin{equation}\label{eLinPolBasDec}
\tilde{T}_{ij}(\mathbf{k},t)=\tilde{T}_+(\mathbf{k},t) \,e^+_{ij}(\hat{{\bf k}}) + \tilde{T}_{\times}(\mathbf{k},t)\,e^{\times}_{ij}(\hat{{\bf k}}).
\end{equation}
The projection operators of Eq.~\eqref{eProjOps} can be written in this basis as
\begin{eqnarray}\label{eProjOpsLin}
	\mathcal{S}_{ijlm}(\hat{\mathbf{k}})&\equiv \frac{1}{2}\left[e^+_{ij} e^+_{lm}(\hat{\mathbf{k}}) + e^{\times}_{ij} e^{\times}_{lm}(\hat{\mathbf{k}})\right],\\
	\mathcal{A}_{ijlm}(\hat{\mathbf{k}})&\equiv \frac{1}{2}\left[e^{\times}_{ij} e^+_{lm}(\hat{\mathbf{k}}) - e^+_{ij} e^{\times}_{lm}(\hat{\mathbf{k}})\right], 
\end{eqnarray} 
and the degree of polarization in of the source can be expressed with
Eq.~\eqref{eDegPolSource}, where in terms of the components in the linear
polarization basis, 
\begin{eqnarray}\label{eDegPolSource2}
T^\mathcal{S}(k)&=\frac{1}{w^2}\int_{\Omega_k} d\Omega_k \, k^2\,\int d^3\mathbf{k}'\left<\tilde{T}^*_+(\mathbf{k})\tilde{T}_+(\mathbf{k}')+\tilde{T}^*_{\times}(\mathbf{k})\tilde{T}_{\times}(\mathbf{k}')\right>, \\
T^\mathcal{A}(k)&=\frac{1}{w^2}\int_{\Omega_k} d\Omega_k \, k^2\,\int d^3\mathbf{k}'\left<\tilde{T}^*_{\times}(\mathbf{k})\tilde{T}_+(\mathbf{k}')-\tilde{T}^*_+(\mathbf{k})\tilde{T}_{\times}(\mathbf{k}')\right>, 
\end{eqnarray} 
where the explicit $t$-dependence is dropped as in Eq. \eqref{eDegPolSourceCirc}. 
Similarly, the spectral energy densities of GWs is written in this basis as 
(again dropping explicit $t$-dependence)
\begin{eqnarray}\label{eEandHGwSpecsLin} 
E_{GW}(k)&=\frac{1}{(2\pi)^6}\frac{1}{16\pi G}\int_{\Omega_k} d\Omega_k \, k^2\,\int d^3\mathbf{k}' \nonumber\\
&\times\left<\dot{\tilde{h}}^*_+(\mathbf{k})\dot{\tilde{h}}_+(\mathbf{k}')+\dot{\tilde{h}}^*_{\times}(\mathbf{k})\dot{\tilde{h}}_{\times}(\mathbf{k}')\right>,\\
H_{GW}(k)&=\frac{1}{(2\pi)^6}\frac{1}{16\pi G}\int_{\Omega_k} d\Omega_k \, k^2\,\int d^3\mathbf{k}' \nonumber\\
&\times\left<\dot{\tilde{h}}^*_{\times}(\mathbf{k})\dot{\tilde{h}}_+(\mathbf{k}')-\dot{\tilde{h}}^*_+(\mathbf{k})\dot{\tilde{h}}_{\times}(\mathbf{k}')\right>. 
\end{eqnarray}

\section{Polarization of the $S$ and $V$ components of the stress tensor}
\label{AppPolSV}

One can define a four-dimensional power spectrum analogous to that in
Eq.~\eqref{eHijlmKandT} from the scalar and vector components of the general
stress tensor $\lambda_{ij}$ of Eq.~\eqref{eSVTMain}.
Indeed using Eq.~\eqref{eLambSVTKSpace}, we can write

\begin{eqnarray}\label{eHijlmKandT-Sca}
&\frac{1}{w^2}\left<k_{\langle i}k_{j\rangle}\tilde{\lambda}^{*S}(\mathbf{k},t)\cdot k'_{\langle l}k'_{m\rangle}\tilde{\lambda}^S(\mathbf{k}',t+\tau)\right> \nonumber\\
&\equiv(2\pi)^3\delta^3(\mathbf{k}-\mathbf{k}')H^S_{ijlm}(\mathbf{k},t,\tau), 
\end{eqnarray}
and 
\begin{eqnarray}\label{eHijlmKandT-Vec}
&\frac{1}{w^2}\left<k_{(i}\tilde{\lambda}^{*V}_{j)}(\mathbf{k},t)\cdot k'_{(l}\tilde{\lambda}^V_{m)}(\mathbf{k}',t+\tau)\right> \nonumber\\
&\equiv(2\pi)^3\delta^3(\mathbf{k}-\mathbf{k}')H^V_{ijlm}(\mathbf{k},t,\tau).
\end{eqnarray}
One can define antisymmetric analogs of these,
\begin{equation}\label{eHijlmKandT-ScaVecAntis}
\mathscr{H}^{S,V}_{ijlm}(\mathbf{k},t,\tau)\equiv -\epsilon_{ipq}\hat{k}_p H^{S,V}_{qjlm}(\mathbf{k},t,\tau),
\end{equation}
with respective ``degrees of polarization'' defined as
\begin{equation}\label{eDegPolSAndV}
\mathcal{P}^{S,V}_T(k,t)=\frac{\ds{\int_{\Omega} d\Omega \, k^2\,\mathscr{H}^{S,V}_{ijij}(\mathbf{k},t,0)}}{\ds{\int_{\Omega} d\Omega \, k^2\,H^{S,V}_{ijij}(\mathbf{k},t,0)}}.
\end{equation}
It follows in a straightforward manner from Eq.~\eqref{eHijlmKandT-Sca} that
$\mathscr{H}^S_{ijij}(\mathbf{k},t,0)=0$, and there is no polarization associated
with the scalar part of the stress tensor, as would be intuitively expected.
However, $\mathscr{H}^V_{ijij}(\mathbf{k},t,0)$ does not necessarily vanish, and
one can associate a degree of polarization with the vector component of the source.


\section*{References}
\begin{thebibliography}{99}

\bibitem{Maggiore:2018sht}
M.~Maggiore,
``Gravitational Waves. Vol. 2: Astrophysics and Cosmology,''
Oxford University Press, Oxford, UK (2018).

\bibitem{Durrer:1997ta}
R.~Durrer and T.~Kahniashvili,
``CMB anisotropies caused by gravitational waves: A Parameter study,''
Helv. Phys. Acta \textbf{71}, 445 (1998).

\bibitem{Weinberg:2003ur}
S.~Weinberg,
``Damping of tensor modes in cosmology,''
Phys. Rev. D \textbf{69}, 023503 (2004).

\bibitem{Mangilli:2008bw}
A.~Mangilli, N.~Bartolo, S.~Matarrese and A.~Riotto,
``The impact of cosmic neutrinos on the gravitational-wave background,''
Phys. Rev. D \textbf{78}, 083517 (2008). 

\bibitem{Lattanzi:2010gn}
M.~Lattanzi, R.~Benini and G.~Montani,
``A possible signature of cosmic neutrino decoupling in the nHz region of the spectrum of primordial gravitational waves,''
Class. Quant. Grav. \textbf{27}, 194008 (2010).

\bibitem{Stefanek:2012hj}
B.~A.~Stefanek and W.~W.~Repko,
``Analytic description of the damping of gravitational waves by free streaming neutrinos,''
Phys. Rev. D \textbf{88},  083536 (2013).

\bibitem{Liu:2015psa}
X.~J.~Liu, W.~Zhao, Y.~Zhang and Z.~H.~Zhu,
``Detecting Relic Gravitational Waves by Pulsar Timing Arrays: Effects of Cosmic Phase Transitions and Relativistic Free-Streaming Gases,''
Phys. Rev. D \textbf{93},  024031 (2016).

\bibitem{Saikawa:2018rcs}
K.~Saikawa and S.~Shirai,
``Primordial gravitational waves, precisely: The role of thermodynamics in the Standard Model,''
JCAP \textbf{05}, 035 (2018).

\bibitem{Deryagin:1987ab}
D. V. Deryagin, D. Y. Grigoriev, V. A. Rubakov, and M. V. Sazhin,
``Generation of gravitational waves by the anisotropic phases in the early universe,''
Mon. Not. Roy. Astron. Soc. \textbf{229}, 357 (1987).

\bibitem{Kahniashvili:2020jgm}
T.~Kahniashvili, A.~Brandenburg, G.~Gogoberidze, S.~Mandal and A.~Roper Pol,
``Circular Polarization of Gravitational Waves from Early-Universe Helical Turbulence,''
Phys. Rev. Res., \textbf{3}, 013193 (2021).

\bibitem{Witten:1984rs}
E.~Witten,
``Cosmic Separation of Phases,''
Phys. Rev. D \textbf{30}, 272 (1984).

\bibitem{Hogan:1986qda}
C.~J.~Hogan,
``Gravitational radiation from cosmological phase transitions,''
Mon. Not. Roy. Astron. Soc. \textbf{218}, 629 (1986).

\bibitem{Kamionkowski:1993fg}
  M.~Kamionkowski, A.~Kosowsky and M.~S.~Turner,
  ``Gravitational radiation from first order phase transitions,''
  Phys.\ Rev.\ D {\bf 49}, 2837 (1994).

\bibitem{Barausse:2020rsu}
E.~Barausse, E.~Berti, T.~Hertog, S.~A.~Hughes, P.~Jetzer, P.~Pani, T.~P.~Sotiriou, N.~Tamanini, H.~Witek and K.~Yagi, \textit{et al.}
``Prospects for Fundamental Physics with LISA,''
Gen. Rel. Grav. \textbf{52},  81 (2020).

\bibitem{Dimopoulos:2007cj}
S.~Dimopoulos, P.~W.~Graham, J.~M.~Hogan, M.~A.~Kasevich and S.~Rajendran,
``Gravitational Wave Detection with Atom Interferometry,''
Phys. Lett. B \textbf{678}, 37 (2009).

\bibitem{Caprini:2010xv}
C.~Caprini, R.~Durrer and X.~Siemens,
``Detection of gravitational waves from the QCD phase transition with pulsar timing arrays,''
Phys. Rev. D \textbf{82}, 063511 (2010).

\bibitem{Ellis:2012in}
J.~A.~Ellis, F.~A.~Jenet and M.~A.~McLaughlin,
``Practical Methods for Continuous Gravitational Wave Detection using Pulsar Timing Data,''
Astrophys. J. \textbf{753}, 96 (2012).

\bibitem{Capozziello:2018qjs}
S.~Capozziello, M.~Khodadi and G.~Lambiase,
``The quark chemical potential of QCD phase transition and the stochastic background of gravitational waves,''
Phys. Lett. B \textbf{789}, 626 (2019).

\bibitem{Arzoumanian:2020vkk}
Z.~Arzoumanian \textit{et al.} [NANOGrav],
``The NANOGrav 12.5 yr Data Set: Search for an Isotropic Stochastic Gravitational-wave Background,''
Astrophys. J. Lett. \textbf{905},  L34 (2020).

\bibitem{Ratzinger:2020koh}
W.~Ratzinger and P.~Schwaller,
``Whispers from the dark side: Confronting light new physics with NANOGrav data,''
SciPost Phys. \textbf{10}  047 (2021).

\bibitem{Neronov:2020qrl}
A.~Neronov, A.~Roper Pol, C.~Caprini and D.~Semikoz,
``NANOGrav signal from MHD turbulence at QCD phase transition in the early universe,''
Phys. Rev. D \textbf{103}, L041302 (2021).

\bibitem{Abe:2020sqb}
K.~T.~Abe, Y.~Tada and I.~Ueda,
``Induced gravitational waves as a cosmological probe of the sound speed during the QCD phase transition,''
[arXiv:2010.06193 [astro-ph.CO]].

\bibitem{Kitajima:2020rpm}
N.~Kitajima, J.~Soda and Y.~Urakawa,
``Nano-Hz gravitational wave signature from axion dark matter,''
%[arXiv:2010.10990 [astro-ph.CO]].
Phys. Rev. Lett. \textbf{126}  121301 (2021).

\bibitem{Ramberg:2020oct}
N.~Ramberg and L.~Visinelli,
``The QCD Axion and Gravitational Waves in light of NANOGrav results,''
Phys. Rev. D \textbf{103},  063031 (2021).
%[arXiv:2012.06882 [astro-ph.CO]].

\bibitem{Li:2021qer}
S.~L.~Li, L.~Shao, P.~Wu and H.~Yu,
``NANOGrav Signal from First-Order Confinement/Deconfinement Phase Transition in Different QCD Matters,''
[arXiv:2101.08012 [astro-ph.CO]].

\bibitem{Gorghetto:2021fsn}
M.~Gorghetto, E.~Hardy and H.~Nicolaescu,
``Observing Invisible Axions with Gravitational Waves,''
[arXiv:2101.11007 [hep-ph]].

\bibitem{Brandenburg:2021tmp}
A.~Brandenburg, E.~Clarke, Y.~He and T.~Kahniashvili,
``Can we observe the QCD phase transition-generated gravitational waves through pulsar timing arrays?,''
[arXiv:2102.12428 [astro-ph.CO]].

\bibitem{He:2021bqm}
Y.~He, A.~Brandenburg and A.~Sinha,
``Spectrum of turbulence-sourced gravitational waves as a constraint on graviton mass,''
JCAP, in press (2021)
[arXiv:2104.03192 [gr-qc]].

\bibitem{Garcia-Bellido:2021zgu}
J.~Garcia-Bellido, H.~Murayama and G.~White,
``Exploring the Early Universe with Gaia and THEIA,''
[arXiv:2104.04778 [hep-ph]].

\bibitem{Sesana:2012ak}
A.~Sesana,
``Systematic investigation of the expected gravitational wave signal from supermassive black hole binaries in the pulsar timing band,''
Mon. Not. Roy. Astron. Soc. \textbf{433}, 1 (2013).

\bibitem{Middleton:2020asl}
H.~Middleton, A.~Sesana, S.~Chen, A.~Vecchio, W.~Del Pozzo and P.~A.~Rosado,
``Massive black hole binary systems and the NANOGrav 12.5 year results,''
Mon. Not. Roy. Astron. Soc. Lett. \textbf{502}, 1 (2021).

\bibitem{Caprini:2018mtu}
C.~Caprini and D.~G.~Figueroa,
``Cosmological Backgrounds of Gravitational Waves,''
Class. Quant. Grav. \textbf{35},  163001 (2018).

\bibitem{Durrer:1999bk}
R.~Durrer, P.~G.~Ferreira and T.~Kahniashvili,
``Tensor microwave anisotropies from a stochastic magnetic field,''
Phys. Rev. D \textbf{61}, 043001 (2000).

\bibitem{Kosowsky:2001xp}
A.~Kosowsky, A.~Mack and T.~Kahniashvili,
``Gravitational radiation from cosmological turbulence,''
Phys. Rev. D \textbf{66}, 024030 (2002).

\bibitem{Melatos:2009mz}
A.~Melatos and C.~Peralta,
``Gravitational Radiation from Hydrodynamic Turbulence in a Differentially Rotating Neutron Star,''
Astrophys. J. \textbf{709}, 77 (2010).

\bibitem{Lasky:2015uia}
P.~D.~Lasky,
``Gravitational Waves from Neutron Stars: A Review,''
Publ. Astron. Soc. Austral. \textbf{32}, e034 (2015).

\bibitem{Neronov:1900zz}
  A.~Neronov and I.~Vovk,
  ``Evidence for strong extragalactic magnetic fields from Fermi observations of TeV blazars,''
  Science {\bf 328}, 73 (2010).
  
  \bibitem{Archambault:2017hvo}
  S.~Archambault,  
  {\it et al.}
  [VERITAS Collaboration],
  ``Search for Magnetically Broadened Cascade Emission From Blazars with VERITAS,''
  Astrophys.\ J.\  {\bf 835},  288 (2017).
  
  \bibitem{Biteau:2018tmv}
  M.~Ackermann, {\it et al.} 
  [Fermi-LAT Collaboration],
  ``The Search for Spatial Extension in High-latitude Sources Detected by the $Fermi$ Large Area Telescope,''
  Astrophys.\ J.\ Suppl.\  {\bf 237}, 32 (2018).
  
  \bibitem{Arlen:2012iy}
  T.~C.~Arlen, V.~V.~Vassiliev, T.~Weisgarber, S.~P.~Wakely and S.~Y.~Shafi,
  ``Intergalactic Magnetic Fields and Gamma Ray Observations of Extreme TeV Blazars,''
  Astrophys.\ J.\  {\bf 796}, 18 (2014).
  
\bibitem{Broderick:2018nqf}
  A.~E.~Broderick, P.~Tiede, P.~Chang, A.~Lamberts, C.~Pfrommer, E.~Puchwein, M.~Shalaby and M.~Werhahn,
  ``Missing Gamma-ray Halos and the Need for New Physics in the Gamma-ray Sky,''
  Astrophys.\ J.\  {\bf 868},  87 (2018).

\bibitem{AlvesBatista:2019ipr}
  R.~Alves Batista, A.~Saveliev and E.~M.~de Gouveia Dal Pino,
  ``The Impact of Plasma Instabilities on the Spectra of TeV Blazars,''
  Mon.\ Not.\ Roy.\ Astron.\ Soc.\  {\bf 489},  3836 (2019).
  
\bibitem{Ahonen:1996nq}
J.~Ahonen and K.~Enqvist,
``Electrical conductivity in the early universe,''
Phys. Lett. B \textbf{382}, 40 (1996).

\bibitem{Brandenburg:1996fc}
A.~Brandenburg, K.~Enqvist and P.~Olesen,
``Large scale magnetic fields from hydromagnetic turbulence in the very early universe,''
Phys. Rev. D \textbf{54}, 1291 (1996).

\bibitem{Brandenburg:2017rnt}
A.~Brandenburg, T.~Kahniashvili, S.~Mandal, A.~Roper~Pol, A.~G.~Tevzadze and T.~Vachaspati,
``Dynamo effect in decaying helical turbulence,''
Phys. Rev. Fluids. \textbf{4}, 024608 (2019).

\bibitem{Kahniashvili:2005qi}
T.~Kahniashvili, G.~Gogoberidze and B.~Ratra,
``Polarized cosmological gravitational waves from primordial helical turbulence,''
Phys. Rev. Lett. \textbf{95}, 151301 (2005).

\bibitem{Kuzmin:1985mm}
V.~A.~Kuzmin, V.~A.~Rubakov and M.~E.~Shaposhnikov,
``On the Anomalous Electroweak Baryon Number Nonconservation in the Early Universe,''
Phys. Lett. B \textbf{155}, 36 (1985).

\bibitem{Shaposhnikov:1986jp}
M.~E.~Shaposhnikov,
``Possible Appearance of the Baryon Asymmetry of the Universe in an Electroweak Theory,''
JETP Lett. \textbf{44}, 465 (1986). 

\bibitem{Cohen:1990py}
A.~G.~Cohen, D.~B.~Kaplan and A.~E.~Nelson,
``Weak Scale Baryogenesis,''
Phys. Lett. B \textbf{245}, 561 (1990).

\bibitem{Cohen:1990it}
A.~G.~Cohen, D.~B.~Kaplan and A.~E.~Nelson,
``Baryogenesis at the weak phase transition,''
Nucl. Phys. B \textbf{349}, 727 (1991).

\bibitem{Morrissey:2012db}
D.~E.~Morrissey and M.~J.~Ramsey-Musolf,
``Electroweak baryogenesis,''
New J. Phys. \textbf{14}, 125003 (2012).

\bibitem{Hindmarsh:2013xza}
M.~Hindmarsh, S.~J.~Huber, K.~Rummukainen and D.~J.~Weir,
``Gravitational waves from the sound of a first order phase transition,''
Phys. Rev. Lett. \textbf{112}, 041301 (2014).

\bibitem{Hindmarsh:2019phv}
M.~Hindmarsh and M.~Hijazi,
``Gravitational waves from first order cosmological phase transitions in the Sound Shell Model,''
JCAP \textbf{12}, 062 (2019).

\bibitem{Christensson:2000sp}
M.~Christensson, M.~Hindmarsh and A.~Brandenburg,
``Inverse cascade in decaying 3-D magnetohydrodynamic turbulence,''
Phys. Rev. E \textbf{64}, 056405 (2001).

\bibitem{Kahniashvili:2010gp}
T.~Kahniashvili, A.~Brandenburg, A.~G.~Tevzadze and B.~Ratra,
``Numerical simulations of the decay of primordial magnetic turbulence,''
Phys. Rev. D \textbf{81}, 123002 (2010).

\bibitem{Vachaspati:2020blt}
T.~Vachaspati,
``Progress on Cosmological Magnetic Fields,''
[arXiv:2010.10525 [astro-ph.CO]].

\bibitem{Lifshitz:1945du} 
E.~Lifshitz,
``Republication of: On the gravitational stability of the expanding universe,''
J.\ Phys.\ (USSR) {\bf 10},  116 (1946)
[Gen.\ Rel.\ Grav.\  {\bf 49}, 18 (2017)].

\bibitem{Grishchuk:1974ny} 
L.~P.~Grishchuk,
``Amplification of gravitational waves in an isotropic universe,''
Sov.\ Phys.\ JETP {\bf 40}, 409 (1975)
[Zh.\ Eksp.\ Teor.\ Fiz.\  {\bf 67}, 825 (1974)].

\bibitem{Mukhanov:1990me} 
V.~F.~Mukhanov, H.~A.~Feldman and R.~H.~Brandenberger,
``Theory of cosmological perturbations. Part 1. Classical perturbations. Part 2. Quantum theory of perturbations. Part 3. Extensions,''
Phys.\ Rept.\  {\bf 215}, 203 (1992).

\bibitem{Durrer:1997ep}
R.~Durrer and M.~Kunz,
``Cosmic microwave background anisotropies from scaling seeds: Generic properties of the correlation functions,''
Phys. Rev. D \textbf{57}, R3199 (1998).

\bibitem{Grasso:1996kk}
D.~Grasso and H.~R.~Rubinstein,
``Revisiting nucleosynthesis constraints on primordial magnetic fields,''
Phys. Lett. B \textbf{379}, 73 (1996).

\bibitem{Zrake:2014mta}
J.~Zrake,
``Inverse cascade of non-helical magnetic turbulence in a relativistic fluid,''
Astrophys. J. Lett. \textbf{794},  L26 (2014).
%doi:10.1088/2041-8205/794/2/L26
%[arXiv:1407.5626 [astro-ph.HE]].


\bibitem{Brandenburg:2014mwa}
A.~Brandenburg, T.~Kahniashvili and A.~G.~Tevzadze,
``Nonhelical inverse transfer of a decaying turbulent magnetic field,''
Phys. Rev. Lett. \textbf{114}, 075001 (2015).
%doi:10.1103/PhysRevLett.114.075001
%[arXiv:1404.2238 [astro-ph.CO]].

\bibitem{Brandenburg:2009tf}
A.~Brandenburg and A.~Nordlund,
``Astrophysical turbulence modeling,''
Rept. Prog. Phys. \textbf{74}, 046901 (2011).
%doi:10.1088/0034-4885/74/4/046901
%[arXiv:0912.1340 [astro-ph.SR]].

\bibitem{Brandenburg:2016odr}
A.~Brandenburg and T.~Kahniashvili,
``Classes of hydrodynamic and magnetohydrodynamic turbulent decay,''
Phys. Rev. Lett. \textbf{118}, 055102 (2017).

\bibitem{Pol:2019yex}
  A.~Roper~Pol, S.~Mandal, A.~Brandenburg, T.~Kahniashvili and A.~Kosowsky,
  ``Numerical Simulations of Gravitational Waves from Early-Universe Turbulence,''
  Phys. Rev. D {\bf 102}, 083512 (2020).

\bibitem{Miniati:2017kah}
F.~Miniati, G.~Gregori, B.~Reville and S.~Sarkar,
``Axion-Driven Cosmic Magnetogenesis during the QCD Crossover,''
Phys. Rev. Lett. \textbf{121} (2018) no.2, 021301.

\bibitem{Kahniashvili:2012vt}
  T.~Kahniashvili, A.~Brandenburg, L.~Campanelli, B.~Ratra and A.~G.~Tevzadze,
  ``Evolution of inflation-generated magnetic field through phase transitions,''
  Phys.\ Rev.\ D {\bf 86}, 103005 (2012).
  %[arXiv:1206.2428 [astro-ph.CO]].
 
\bibitem{Mee:2006mq}
A.~J.~Mee and A.~Brandenburg,
``Numerical turbulence forced through localized random expansion waves,''
Mon. Not. Roy. Astron. Soc. \textbf{370}, 415 (2006).

\bibitem{KP73}
B. B. Kadomtsev and V. I. Petviashili,
``Acoustic turbulence,''
Sov. Phys. Dokl. \textbf{18}, 115 (1973).

\bibitem{Gogoberidze:2007an}
  G.~Gogoberidze, T.~Kahniashvili and A.~Kosowsky,
  ``The Spectrum of Gravitational Radiation from Primordial Turbulence,''
  Phys.\ Rev.\ D {\bf 76}, 083002 (2007).
  
\bibitem{Kahniashvili:2008er}
T.~Kahniashvili, G.~Gogoberidze and B.~Ratra,
``Gravitational Radiation from Primordial Helical MHD Turbulence,''
Phys. Rev. Lett. \textbf{100}, 231301 (2008).

\bibitem{Brandenburg:2019uzj}
A.~Brandenburg and S.~Boldyrev,
``The turbulent stress spectrum in the inertial and subinertial ranges,''
Astrophys. J. \textbf{892}, 80 (2020). 
%[arXiv:1912.07499 [astro-ph.CO]].

\bibitem{Pol:2018pao}
  A.~Roper Pol, A.~Brandenburg, T.~Kahniashvili, A.~Kosowsky and S.~Mandal,
  ``The timestep constraint in solving the gravitational wave equations sourced by hydromagnetic turbulence,''
  Geophys. Astrophys. Fluid Dynamics \textbf{114}, 130 (2020).

\bibitem{Caprini:2003vc}
  C.~Caprini, R.~Durrer and T.~Kahniashvili,
  ``The Cosmic microwave background and helical magnetic fields: The Tensor mode,''
  Phys.\ Rev.\ D {\bf 69}, 063006 (2004).
  
\bibitem{monin-yaglom}
A. S. Monin and A. M. Yaglom. {\it Statistical
Fluid mechanics
mechanics of turbulence}, vol. 2. MIT press, Cambridge, 1971.

\bibitem{Falk94}
G. Falkovich,
``Bottleneck phenomenon in developed turbulence,''
Phys. Fluids \textbf{6}, 1411 (1994).

\bibitem{Stein:1967}
R. F. Stein,
``Generation of Acoustic and Gravity Waves by Turbulence in an Isothermal Stratified Atmosphere,''
Solar Physics \textbf{2}, 385 (1967).

\bibitem{Dobler:2003xh}
W.~Dobler, N.~E.~L.~Haugen, T.~A.~Yousef and A.~Brandenburg,
``The bottleneck effect in three - dimensional turbulence simulations,''
Phys. Rev. E \textbf{68}, 026304 (2003).
%[arXiv:astro-ph/0303324 [astro-ph]].

\bibitem{PC} 
Pencil Code Collaboration:
A. Brandenburg, A. Johansen, P. A. Bourdin, W. Dobler, W. Lyra, M. Rheinhardt, S. Bingert, N. E. L. Haugen, A. Mee, F. Gent, N. Babkovskaia, C.-C. Yang, T. Heinemann, B. Dintrans, D. Mitra, S. Candelaresi, J. Warnecke, P. J. K\"apyl\"a, A. Schreiber, P. Chatterjee, M. J. K\"apyl\"a, X.-Y. Li, J. Kr\"uger, J. R. Aarnes, G. R. Sarson, J. S. Oishi, J. Schober, R. Plasson, C. Sandin, E. Karchniwy, L. F. S. Rodrigues, A. Hubbard, G. Guerrero, A. Snodin, I. R. Losada, J. Pekkil\"a, and C. Qian\yjour{2021}{Journal of Open Source Software}{6}{2807}
{The Pencil Code, a modular MPI code for partial differential equations and particles: multipurpose and multiuser-maintained}

\bibitem{DATA}
T. Kahniashvili, A. Brandenburg, G. Gogoberidze, S. Mandal, and A. Roper~Pol,
Datasets for ``Circular Polarization of Gravitational Waves from Early-Universe Helical Turbulence''
(v2020.11.07), \url{https://doi.org/10.5281/zenodo.4256906}; see also
\url{http://www.nordita.org/~brandenb/projects/CircPol/} for easier access.

\bibitem{DATA2}
A. Brandenburg, E. Clarke, Y. He, T. Kahniashvili,
Datasets for ``Can we observe the QCD phase transition-generated gravitational waves through pulsar timing arrays?''
(v2021.02.24), \url{https://doi.org/10.5281/zenodo.4560423}; see also
\url{http://www.nordita.org/~brandenb/projects/GWfromQCD/} for easier access.

\end{thebibliography}
\end{document}